\documentclass[prx,aps,showpacs,twocolumn]{revtex4}
\usepackage[T1]{fontenc}
\usepackage{fourier}
\usepackage{graphicx}
\usepackage{float}
\usepackage{latexsym,amsmath,amssymb,bm,euscript}
\usepackage{color}
\usepackage{subfigure}
\usepackage{bm}
\usepackage[colorlinks=true,linkcolor=blue,citecolor=blue]{hyperref}
\usepackage{soul}

\begin{document}

\title{Emergent quasi-one-dimensionality in a kagom\'e magnet: A simple route to complexity}
\author{Shou-Shu Gong$^{1}$, Wei Zhu$^2$, Kun Yang$^3$, Oleg A. Starykh$^4$, D. N. Sheng$^2$, and Leon Balents$^5$}
\affiliation{$^{1}$National High Magnetic Field Laboratory, Florida State University, Tallahassee, FL 32310\\
$^{2}$Department of Physics and Astronomy, California State University, Northridge, CA 91330\\
$^{3}$National High Magnetic Field Laboratory and Department of Physics, Florida State University, Tallahassee, FL 32306\\
$^{4}$Department of Physics and Astronomy, University of Utah, Salt Lake City, UT 84112\\
$^{5}$Kavli Institute for Theoretical Physics, University of California, Santa Barbara, CA 93106}

\begin{abstract}
  We study the ground state phase diagram of the quantum spin-$1/2$
  Heisenberg model on the kagom\'{e} lattice with first- ($J_1 < 0$),
  second- ($J_2 < 0$), and third-neighbor interactions ($J_d > 0$) by
  means of analytical low-energy field theory and numerical
  density-matrix renormalization group (DMRG) studies.  The results offer a
  consistent picture of the $J_d$-dominant regime in terms of three
  sets of spin chains weakly coupled by the ferromagnetic inter-chain
  interactions $J_{1,2}$.  When either $J_1$ or $J_2$ is much stronger than the other one, the
  model is found to support one of two cuboctohedral phases, cuboc1
  and cuboc2.  These cuboc states host non-coplanar long-ranged
  magnetic order and possess finite scalar spin chirality. However, in
  the compensated regime $J_1 \simeq J_2$, a valence bond crystal
  phase emerges between the two cuboc phases.  We find excellent
  agreement between an analytical theory based on coupled spin chains
  and unbiased DMRG calculations, including at a very detailed level
  of comparison of the structure of the valence bond crystal state.
  To our knowledge, this is the first such comprehensive understanding
  of a highly frustrated two-dimensional quantum antiferromagnet.  We find no
  evidence of either the one-dimensional gapless spin liquid or the
  chiral spin liquids, which were previously suggested by parton mean
  field theories.
\end{abstract}

\maketitle

\section{Introduction}
\label{sec:introduction}

The kagom\'e lattice, from its humble origin in the hands of
fishermen, now sits at the forefront of the search for exotic quantum
phases of matter such as quantum spin liquids \cite{balents2010, savary2016}.
Compelling numerical evidence shows that, on this lattice, even the
simplest model of magnetism, the Heisenberg spin-$1/2$ Hamiltonian with
up to third neighbor interactions, shows not just one but at least two
of these highly entangled phases \cite{mendels2007, helton2007, vries2009, han2012,
fu2015, han2015, jiang2008, yan2011, depenbrock2012, jiang2012nature, ran2007, iqbal2013, iqbal2014,
messio2012, gong2014kagome, he2014csl, bauer2014, gong2015kagome, hu2015, lauchli2015kagome, yang1993, zhu2015,
he2015, tu2016,zaletel2015, cincio2015}.
The vast majority of studies have naturally focused on the regime in which
antiferromagnetic nearest-neighbor coupling is dominant. Recently, however,
theory and experiment have turned to a different limit, of
dominant antiferromagnetic {\em third-neighbor} interaction $J_d$ (of
a particular type, across the diagonal of the kagom\'e lattice), in
the material kapellasite, Cu$_3$Zn(OH)$_6$Cl$_2$ \cite{colman2008, colman2010,
fak2012, ker2014, janson2008, jeschke2013}.
In this regime, the proposed model, which also has the ferromagnetic $J_1, J_2$
couplings, classically supports interesting non-coplanar ground states with
spins pointing to the corners of a cuboctohedron, a state which leaves
no residual continuous subgroup of SU(2) spin-rotation symmetry unbroken, and possesses spontaneous non-zero scalar spin chirality \cite{messio2011}.  It has been suggested
based on mean-field and variational parton constructions that for $S=1/2$ 
quantum fluctuations may overcome these orders, leading to chiral quantum spin 
liquid ground states \cite{bieri2014}.    

In the large-$J_d$ limit, the kagom\'e lattice separates into
a mesh-like set of three kinds of one-dimensional (1d) chains oriented at
$\pm 120^\circ$ to each other.  We show that such a separation is more
than just a geometric curiosity. It allows us to capture {\em all} the 
low-energy degrees of freedom of the model, including those of the
emergent non-local dimer fluctuations.  A careful analysis of the
residual inter-chain interactions, all of which are represented by the
weak $J_1$ and $J_2$ bonds, offers us a complete understanding of the
phases of this strongly frustrated two-dimensional (2d) spin-$1/2$
Heisenberg model. We find the phase diagram to contain two
cuboctohedral phases, denoted cuboc1 and cuboc2, separated by a region
of valence bond crystal (VBC) order. Deep in the analytical limit with
$|J_1|, |J_2| \ll J_d$, but $J_1/J_2$ arbitrary, we obtain asymptotically
exact results for the phase boundaries between these three phases.
Our analytical approach is nicely complemented by highly accurate
numerical Density Matrix Renormalization Group (DMRG) computations on
kagom\'e cylinders of two different geometries, with a circumference
of up to 12 sites.  Our numerical results strongly support the phase
diagram consisting of two cuboc and VBC phases as predicted
analytically.  We find {\em no} evidence of the suggested chiral spin
liquid states, either analytically or numerically, which puts the
validity of the parton approximation for this problem \cite{bieri2014}
into question.  At small $J_1,J_2$, this agrees with a pseudofermion functional
renormalization group calculation \cite{iqbal2015}, which however finds a spin liquid
state for intermediate $J_1$, which we do not observe.
Profound implications of our findings, both for the minimal
theoretical model of kapellasite and for the physics of the real
material, are discussed in the concluding section of the paper.

\begin{figure}[h!]
  \centering
  \includegraphics[width=0.75\columnwidth]{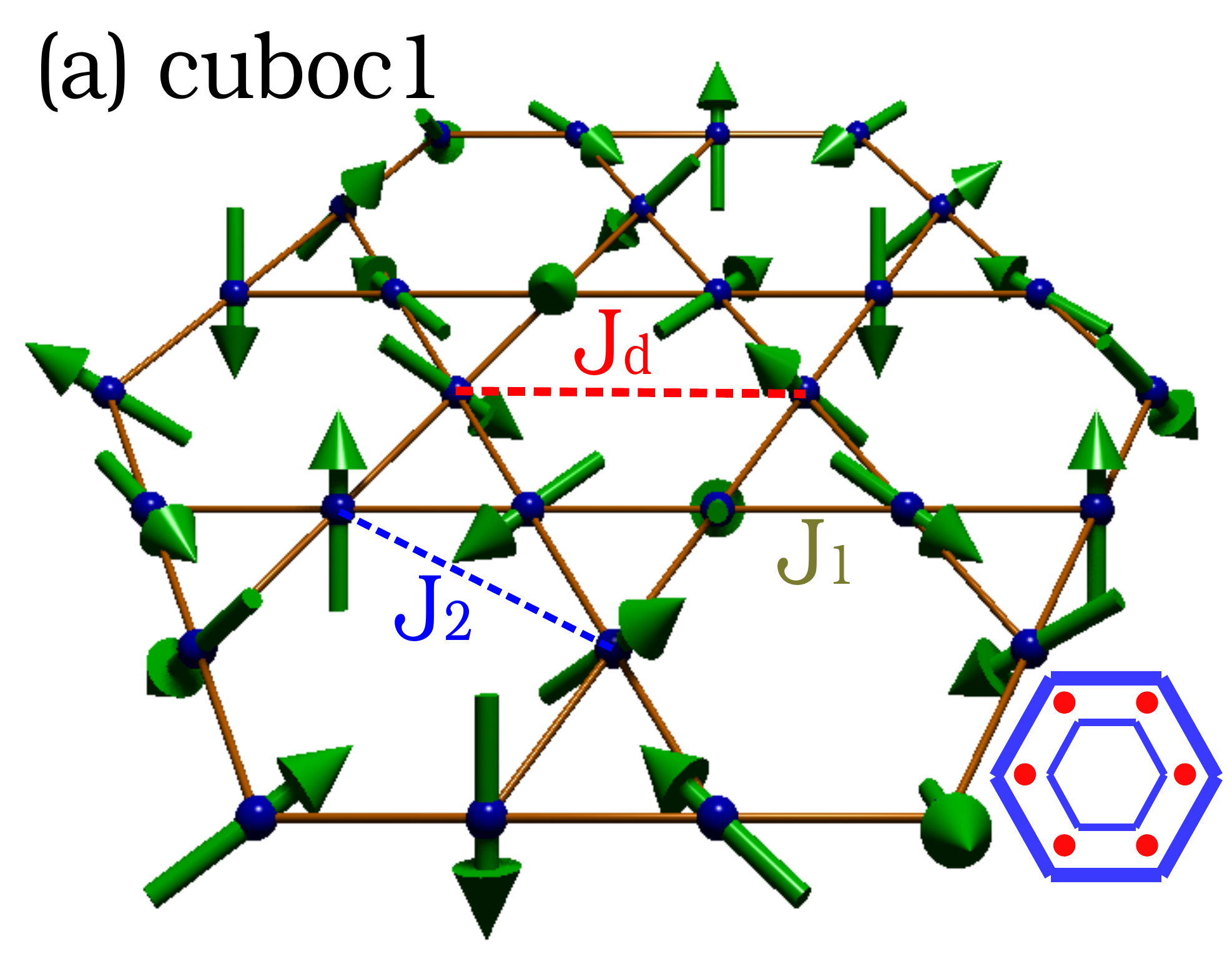}
  \includegraphics[width=0.75\columnwidth]{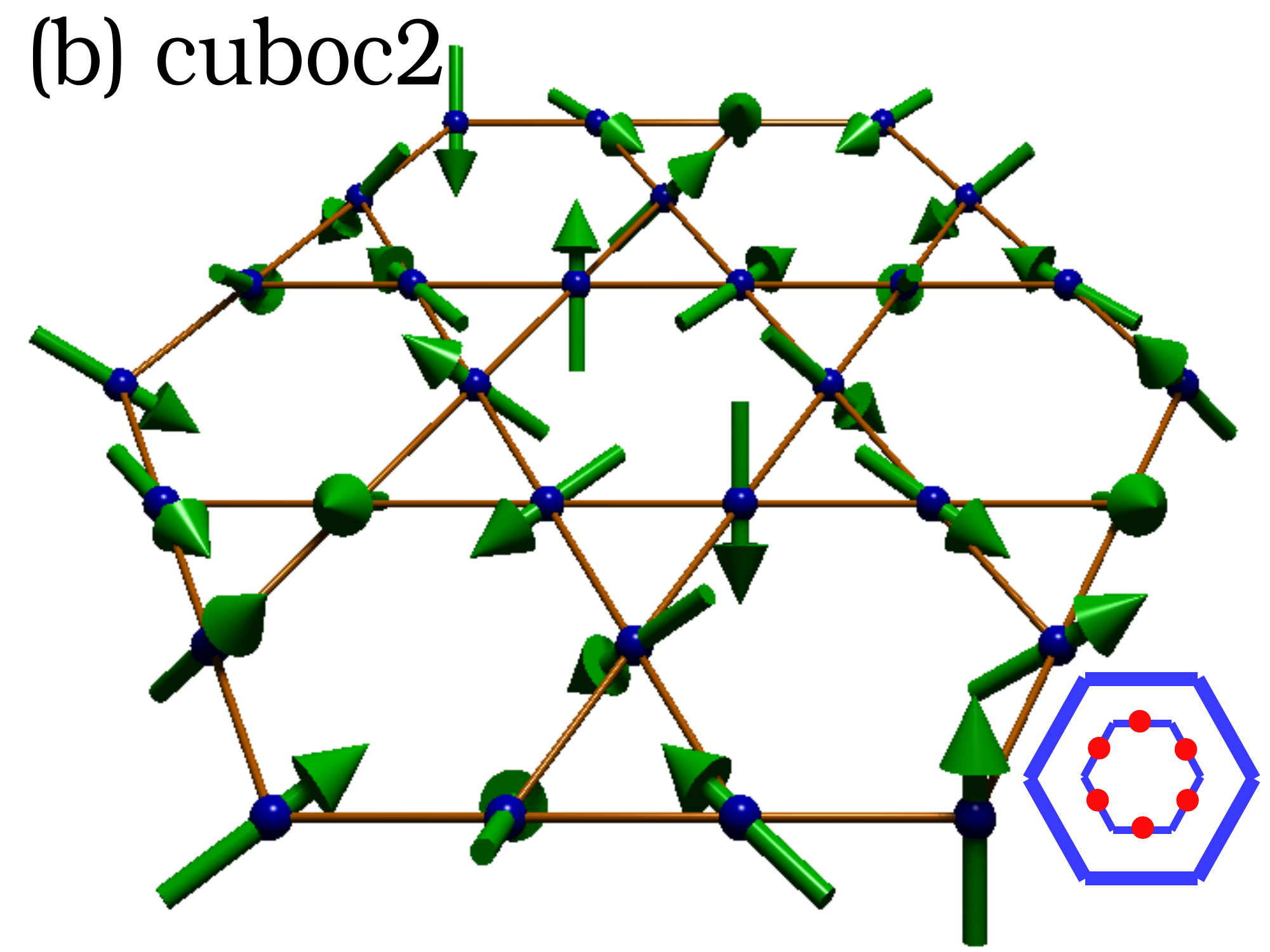}
  \caption{(Color online) Spin configuration in the ordered cuboc
    states on kagom\'{e} lattice.  The arrows indicate the direction
    of static moments.  (a) Cuboc1 state. In this state, the spins on triangles are
    coplanar.  In each hexagon, sets of three consecutive spins are
    non-coplanar.  The $J_1, J_2, J_d$ bonds denote the Heisenberg
    interactions of the Hamiltonian Eq.~\eqref{hamiltonian}.
    (b) Cuboc2 state. In this state, the spins on triangles are non-coplanar
    and those on hexagons are coplanar. The insets indicate the static
    spin structure factors for the cuboc
    states in momentum space with the peaks shown by red dots. The
    smaller hexagon is the Brillouin zone of the kagom\'{e} lattice,
    and the larger one is the extended Brillouin zone of the extended
    triangular lattice by adding a virtual site in the center of each
    hexagon on kagom\'{e} lattice.}
  \label{fig:cuboc}
\end{figure}

\begin{figure}[h!]
  \centering
  \includegraphics[width=1.0\columnwidth]{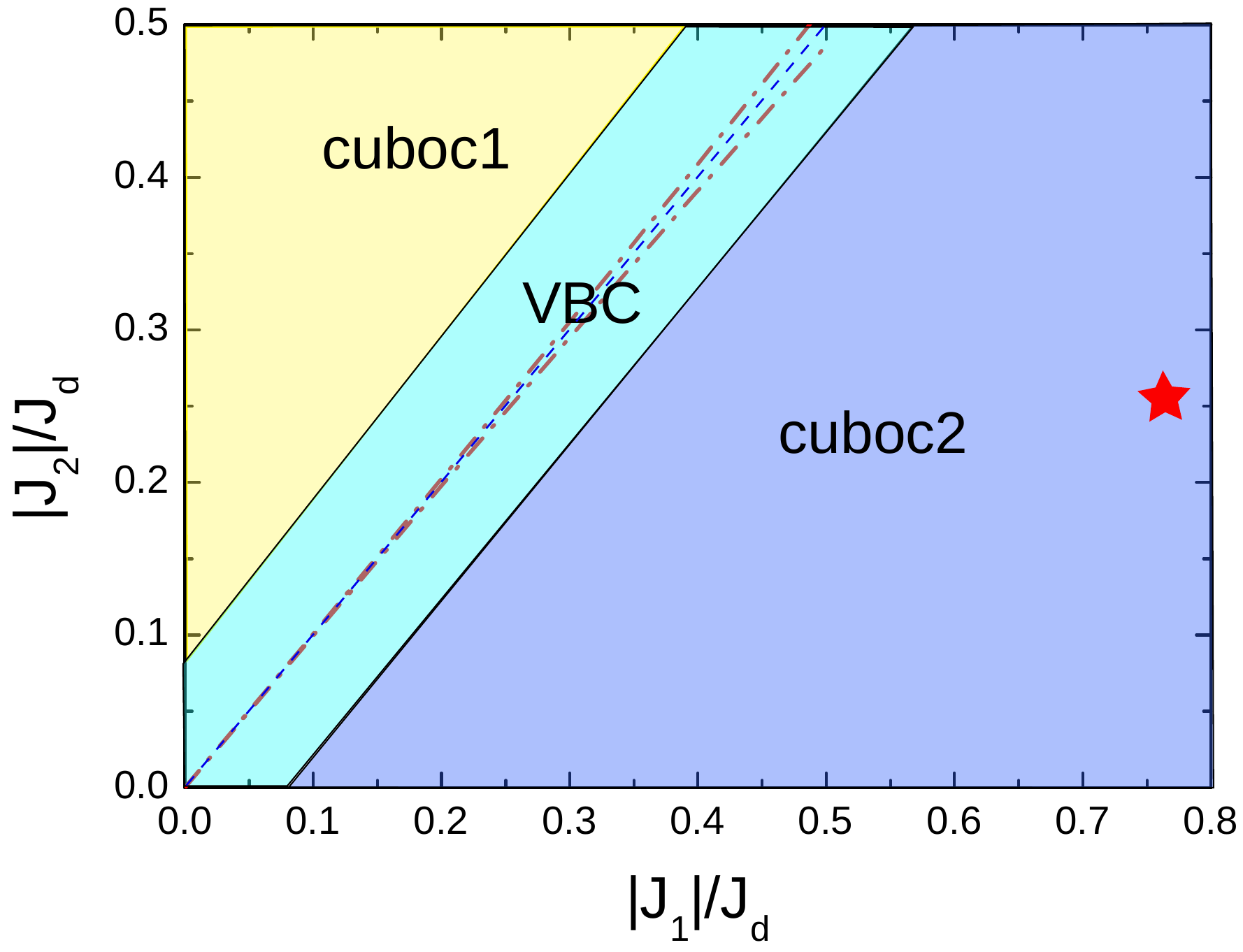}
  \caption{(Color online) Quantum phase diagram of the spin-$1/2$
    $J_1$-$J_2$-$J_d$ kagom\'{e} model.  The short-dashed (blue) lines denote the
    classical phase boundary separating the cuboc1 and cuboc2 phases \cite{bieri2014}. 
    For the spin-$1/2$ Hamiltonian Eq.~\eqref{hamiltonian}, 
    the direct phase transition between the two cuboc phases
    is replaced by the $24$-fold degenerate spontaneously dimerized 
    VBC phase (see Sec.~\ref{sec:compensated-regime} for the analysis).
    The two-dimensional phase boundaries with the cuboc states are shown by the dot-dashed (red) lines (see Eq.~\eqref{eq:32}),
    which form the wedge-like shape of the 2d VBC phase. 
    The DMRG calculations on open cylinders with different geometries 
    (see Sec.~\ref{sec:dmrg-results}) find a wider dimerized region, 
    as indicated by the cyan-colored stripe centered around the compensated line $J_1 = J_2$. The symmetries
    of the dimer order found in DMRG are fully consistent with those of the two-dimensional VBC phase. 
    The enhancement of the VBC order in the cylinder geometry is caused by the strong finite size effects 
    due to open boundary  conditions. The red star represents parameters best describing kapellasite \cite{bernu2013}.}
  \label{fig:phase}
\end{figure}

\section{Analytical treatment}
\label{sec:analytical-treatment}

\subsection{Mapping to coupled chains and scaling operators}
\label{sec:mapp-coupl-chains}

We consider the Heisenberg Hamiltonian,
\begin{equation}
 H = J_1 \sum_{\langle i,j \rangle} {\bm S}_i \cdot \bm{S}_j
  + J_2 \sum_{\langle\langle i,j \rangle\rangle} \bm{S}_i \cdot \bm{S}_j
  + J_d \sum_{\langle\langle\langle i,j \rangle\rangle\rangle_d}
  \bm{S}_i \cdot \bm{S}_j,\label{hamiltonian}
\end{equation}
where the respective interactions are as shown in Fig.~\ref{fig:cuboc}(a).
To make analytic progress, we assume $|J_1|, |J_2| \ll J_d$.  Then the exchanges
$J_1,J_2$ have substantial effects only at low energy, and we can use
the field theory representation in this regime for the disjoint one
dimensional chains generated by $J_d$ alone.  These chains form three
families, oriented at 120$^\circ$ to one another.  Within each family,
we take ${\sf x}$ the coordinate along the chain and ${\sf y}$ to
label the chain itself, with ${\sf x}, {\sf y}$ integers. Taking the
nearest-neighbor lattice spacing to unity, we define primitive lattice
vectors ${\bm a}_1=(2,0)$, $ {\bm a}_2=(-1,\sqrt{3})$, and
${\bm a}_3=-{\bm a}_1-{\bm a}_2$, which connect unit cells.  The spin
on chain of type $q=1,2,3$ with ``chain'' coordinates
${\sf x},{\sf y}$ is located in real space at
${\bm x} = ({\sf x}+\frac{1}{2}){\bm a}_q + {\sf y} {\bm a}_{q+1}$,
where here and in the following we treat $q$ as periodic,
i.e. $q=3+1\equiv 1$.  For an isotropic system, the total number of
sites is $\mathcal{N} = 3 {\sf L}^2$, where ${\sf L}$ is both the
number of sites in a chain (range of ${\sf x}$) and the number of
chains of a single orientation (range of ${\sf y}$).  In this chain
notation, we can rewrite the interchain interactions as
\begin{eqnarray}
  \label{eq:2}
  H' & = & J_1 \sum_{\sf y,y'} \sum_q \big({\bm S}_{q,{\sf y}}(-{\sf y}') \cdot {\bm S}_{q+1,{\sf y}'}({\sf y}+{\sf y}' -1)  \nonumber \\
      &+& {\bm S}_{q,{\sf y}}(-{\sf y}'-1) \cdot{\bm S}_{q+1,{\sf y}'}({\sf y}+{\sf y}')\big) \nonumber \\
      &+& J_2 \sum_{\sf y,y'} \sum_q \big({\bm S}_{q,{\sf y}}(-{\sf y}'-1)\cdot {\bm S}_{q+1,{\sf y}'}({\sf y} + {\sf y}' -1)  \nonumber \\
      &+& {\bm S}_{q,{\sf y}}(-{\sf y}') \cdot {\bm S}_{q+1,{\sf y}'}({\sf y}+{\sf y}')\big),
\end{eqnarray}
where ${\bm S}_{q,{\sf y}}({\sf x})$ is the spin in chain coordinates.

\begin{figure}[h!]
  \centering
  \includegraphics[width=0.5\columnwidth]{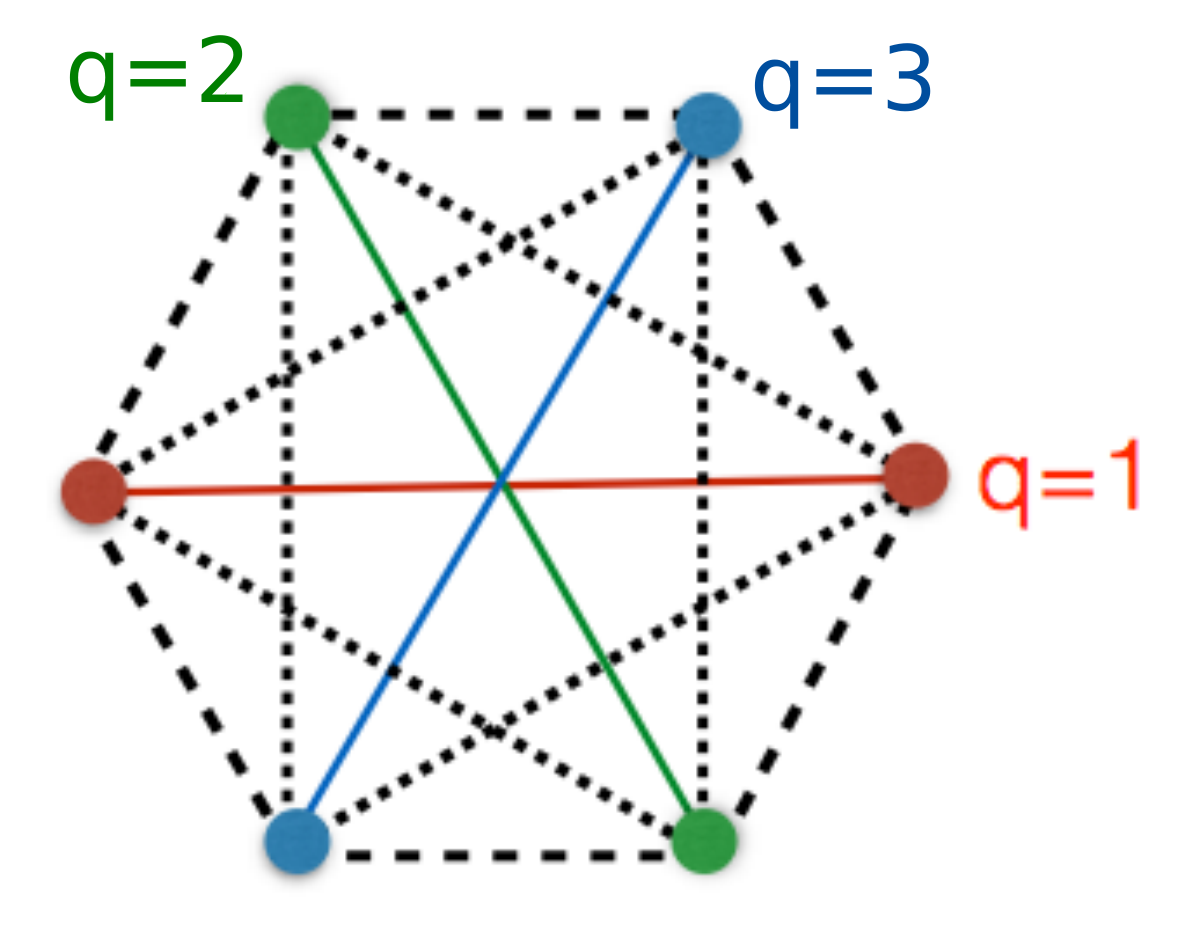}
  \caption{(Color online) An elementary hexagon of the kagom\'e
    lattice. The three types of chains formed by $J_d$ couplings,
    labeled $q=1,2,3$, are shown in
    different colors.  Interchain interactions $J_1$ ($J_2$) are shown
    by dashed (dotted) lines. Observe that for $J_1=J_2$ the
    interchain interaction can be written as a scalar product of sums
    of pairs of spins from chains with different $q$'s, as written in
    Eq.~\eqref{eq:4}. The total interchain Hamiltonian is given by the
    sum over all hexagons of the lattice.}
  \label{fig:hexagon}
\end{figure}

At low energy, each chain, labeled by $q$ and ${\sf y}$, is described by 
a Wess-Zumino-Witten (WZW) SU(2)$_1$ theory, which has primary 
fields ${\bm N}_{q,{\sf y}}$ and $\varepsilon_{q,{\sf y}}$, describing 
staggered magnetization (N\'eel) and staggered dimerization, respectively, 
as well as chiral SU(2) currents ${\bm J}_{q,{\sf y},R}, {\bm J}_{q,{\sf y},L}$.  
The lattice spin operators decompose into
\begin{equation}
  \label{eq:1}
  {\bm S}_{q,{\sf y}}({\sf x}) = (-1)^{\sf x} {\bm N}_{q,{\sf y}}({\sf
    x}) + {\bm M}_{q,{\sf y}}({\sf x}),
\end{equation}
where ${\bm M}= {\bm J}_R + {\bm J}_L$ is the uniform magnetization.
The fields ${\bm N}_{q,{\sf y}}({\sf x})$ and ${\bm M}_{q,{\sf
    y}}({\sf x})$ can be treated as slowly-varying functions of ${\sf
  x}$.
The primary fields have scaling dimension $\Delta=1/2$, and represent
the strongest correlations of Heisenberg chains.  The currents have
larger scaling dimension $\Delta=1$, and so are
less important within interactions than the primary fields.  Hence the
dominant interaction is generically given by using Eq.~\eqref{eq:1}
and keeping the N\'eel fields alone:
\begin{equation}
  \label{eq:3}
  H'_{\rm dom} \sim 2 (J_2-J_1) \sum_q \sum_{\sf y,y'} (-1)^{\sf y} {\bm
    N}_{q,{\sf y}}(-{\sf y}') \cdot {\bm N}_{q+1,{\sf y}'}({\sf
    y}+{\sf y}').
\end{equation}
We observe that $J_1$ and $J_2$ give identical contributions in this
leading approximation, only of opposite sign.  This leads to a
vanishing along the compensated line $J_1=J_2$.  In the vicinity of
this line, otherwise sub-dominant terms will play a role.  At the
lattice level, the compensation is already evident, as we can rewrite
Eq.~\eqref{eq:2} in this case as
\begin{widetext}
  \begin{equation}
    \label{eq:4}
    \left. H'\right|_{J_1=J_2} = J_1  \sum_{\sf y,y'} \sum_q \left(
      {\bm S}_{q,{\sf y}}(-{\sf y}') +  {\bm S}_{q,{\sf y}}(-{\sf
        y}'-1) \right)\cdot  \left(
      {\bm S}_{q+1,{\sf y}'}({\sf y}+{\sf y}') +  {\bm S}_{q+1,{\sf
          y}'}({\sf y}+{\sf
        y}'-1) \right).
  \end{equation}
  This symmetric form can be seen directly from examination of the
  interactions between two chains that cross at a given hexagon,
    see Fig.~\ref{fig:hexagon}.  Now using Eq.~\eqref{eq:1}, the leading scaling
  term becomes
\begin{equation}
  \label{eq:5}
   \left. H'_{\rm sub-dom}\right|_{J_1=J_2} \sim J_1  \sum_{\sf y,y'}
   \sum_q \left(2{\bm M}_{q,{\sf y}}(-{\sf y}') +(-1)^{{\sf y}'} \partial_{\sf x}{\bm N}_{q,{\sf
         y}}(-{\sf y}')\right) \cdot \left(2 {\bm M}_{q+1,{\sf
         y}'}({\sf y}+{\sf y}') + (-1)^{{\sf y}+{\sf y}'}\partial_x{\bm N}_{q+1,{\sf y}'}({\sf
       y}+{\sf y}') \right)
\end{equation}
\end{widetext}
The term in Eq.~\eqref{eq:5} is only important when the leading one in
Eq.~\eqref{eq:3} is nearly zero, so it is legitimate to take
$J_1\approx J_2$ in the former.

\subsection{Two dimensional analysis}
\label{sec:two-dimens-analys}

The leading interactions in Eq.~\eqref{eq:3} and the sub-dominant
corrections in Eq.~\eqref{eq:5} can be analyzed directly for the
two-dimensional infinite system, following methods developed for other
coupled chain systems.  The closest analogy is that for the planar
pyrochlore lattice in Ref.~\onlinecite{starykh2005anisotropic}.

\subsubsection{Magnetically ordered region}
\label{sec:magn-order-regi}

Away from the compensated line, it is sufficient to consider the
dominant term in Eq.~\eqref{eq:3}.  At a na\"ive scaling level, since
this interaction is represented in field theory by a term in the
effective actions $S' = \int \!d\tau\, H'$, and each ${\bm N}$ field
has dimension $\Delta[{\bm N}]=1/2$, the interaction is marginal.
However, as shown in Ref.~\onlinecite{starykh2005anisotropic}, the presence
of a number of such terms proportional to the system volume amplifies
its effect, inducing magnetic order with a strength that is power-law
in the coupling ($J_1-J_2$).  This can be verified by using chain mean
field theory, in which each chain is treated exactly but the
interaction between chains is decoupled in mean field fashion.  It can
be formulated as a variational approximation, with the variational
state the ground state of a fiduciary mean field Hamiltonian, which is
\begin{equation}
  \label{eq:6}
  H_{\rm MF} = H_0 - \sum_{q,{\sf y}} \int \! d{\sf x}\, {\bm
    h}_{q,{\sf y}} \cdot {\bm N}_{q,{\sf y}}({\sf x}),
\end{equation}
where $H_0 = H |_{J_1=J_2=0}$ and ${\bm h}_{q,{\sf y}}$ are effective
staggered fields for each chain, which are treated as variational
parameters.  The variational energy is
\begin{equation}
  \label{eq:7}
  E_{\rm var} = \langle H_0 \rangle-E_0 + \langle H'_{\rm dom}\rangle,
\end{equation}
where the expectation values are taken in the variational state, and
for convenience we subtracted the energy $E_0$ of pure Heisenberg
chains so that the zero of energy is realized at $J_1=J_2=0$.  This
can be simplified using the fact that $\bar{\bm N}_{q,{\sf y}} \equiv \langle {\bm N}_{q,{\sf
    y}}({\sf x})\rangle = \alpha |{\bm h}_{q,{\sf y}}|^{1/3} \hat {\bm h}_{q,{\sf y}}$, where $\alpha$
is a constant, which follows from the spin-rotation symmetry of $H_0$
and the scaling dimension of ${\bm N}$ \cite{Schulz1996}.  Similarly, the mean-field
energy obeys $\langle H_{\rm MF} \rangle = E_0 - \frac{3}{4} \alpha
{\sf L} \sum_{q,{\sf
    y}}  |{\bm h}_{q,{\sf y}}|^{4/3}$.  We can therefore express all
variational energies in terms of $\bar{\bm N}_{q,{\sf y}} $, using
Eq.~\eqref{eq:6} to obtain
\begin{equation}
  \label{eq:8}
  E_{\rm var}= \frac{{\sf L}}{4\alpha^3} \sum_{q,{\sf y}} |\bar{\bm
    N}_{q,{\sf y}}|^4 - \lambda \sum_q \sum_{\sf y,y'} (-1)^{\sf y} \bar{\bm
    N}_{q,{\sf y}}\cdot \bar{\bm N}_{q+1,{\sf y}'},
\end{equation}
where $\lambda = 2(J_1-J_2)$.
The instability to ordering is evident from the fact that
interaction (second) term, which can be made negative by a suitable variational
choice, is quadratic in $\bar{N}$, and hence dominates over the first
term, which is quartic, for small $\bar{N}$.  This implies that
$E_{\rm var}$ always has a non-trivial global minimum, even for
arbitrarily small $\lambda$.  The task is now to find this minimum.

To do so, we assume that $|\bar{\bm N}_{q,{\sf y}}| = \bar{N}$ is
constant for all chains.  Then we minimize the second term over the
orientations
$\hat{\bm N}_{q,{\sf y}}= \bar{\bm N}_{q,{\sf y}}/|\bar{\bm N}_{q,{\sf
    y}}|$, and after this finally minimize over $\bar{N}$.  We can rewrite the variational energy as
\begin{equation}
  \label{eq:9}
  \frac{E_{\rm var}}{\mathcal{N}} = \frac{1}{4\alpha^3} \bar{N}^4 - \frac{\lambda}{3}
  \bar{N}^2 \sum_q {\bm V}_{q} \cdot {\bm W}_{q+1},
\end{equation}
where
\begin{equation}
  \label{eq:10}
  {\bm V}_{q} = \frac{1}{\sl L} \sum_y (-1)^y \hat{\bm N}_{q,{\sf y}},
  \qquad {\bm W}_q = \frac{1}{\sl L} \sum_y \hat{\bm N}_{q,{\sf y}}.
\end{equation}
and the fixed magnitude constraint implies that
\begin{equation}
  \label{eq:11}
  |{\bm V}_q| = |{\bm W}_q|=1, \qquad {\bm V}_q \cdot {\bm W}_q=0.
\end{equation}

Obviously the $\lambda$ term is bounded below by
$-|\lambda|\bar{N}^2$, which is achieved if and only if ${\bm V}_q
\cdot {\bm W}_{q+1} =  {\rm sign} (\lambda)$.   The general solution
of these conditions is
\begin{equation}
  \label{eq:12}
  {\bm V}_q = \hat{\bm e}_{q+1}, \qquad {\bm W}_q =  {\rm sign}(\lambda)
  \hat{\bm e}_{q},
\end{equation}
where $\hat{\bm e}_1,\hat{\bm e}_2, \hat{\bm e}_3$ are three
orthonormal vectors.  Finally minimization gives
\begin{equation}
  \label{eq:13}
  \bar{N} = \sqrt{ 2 \alpha^3 |\lambda|}
\end{equation}
and $E_{\rm var} = - \alpha^3 \lambda^2 {\mathcal{N}}$.
We see that the ordered moment grows as the square root of the
inter-chain couplings, when we are away from the compensated line.

Clearly the solutions in Eq.~\eqref{eq:12} represent non-coplanar
magnetically ordered configurations.  In fact these are exactly the
cuboc1 ($\lambda>0$) and cuboc2 ($\lambda<0$) states expected
classically. On chains of type $q$, the ordered
moments are oriented along the four directions
$\hat{\bm e}_q+\hat{\bm
  e}_{q+1}$,$-\hat{\bm
  e}_q-\hat{\bm e}_{q+1}$,$\hat{\bm
  e}_q-\hat{\bm e}_{q+1}$,$-\hat{\bm
  e}_q+\hat{\bm e}_{q+1}$,
and all spins together define 12 unique sublattices.  Spins along any
given chain are all collinear, and alternate in a normal N\'eel
pattern.  The state also has non-zero scalar chirality, which is of
order $\bar{N}^3$ and hence parametrically smaller than the ordered
moment in the weakly coupled chain limit.  The sign of the scalar
chirality is a discrete order parameter and proportional to the triple
product $\hat{\bm e}_1 \cdot ( \hat{\bm e}_2 \times \hat{\bm e}_3 )$.

 To see this, it is convenient to write ${\bm V}_q=(\hat{\bm N}_q^{(e)} - \hat{\bm N}_q^{(o)})/2$ and
 ${\bm W}_q = (\hat{\bm N}_q^{(e)}+ \hat{\bm N}_q^{(o)})/2$ in terms of
 unit fields on {\it even}, $\hat{\bm N}_q^{(e)} = \hat{\bm N}_{q,\sf{y}=\rm{even}}$, and {\it odd},
 $\hat{\bm N}_q^{(o)} = \hat{\bm N}_{q,{\sf{y}=\rm{odd}}}$, chains of type $q$.
 Then Eq.~\eqref{eq:12} tells that $\hat{\bm N}_q^{(e/o)} = {\rm sign}(\lambda) \hat{\bm e}_{q} \pm  \hat{\bm e}_{q+1}$.
 Therefore, using Eqs.~\eqref{eq:1} and \eqref{eq:13}, our chain mean-field theory predicts for the expectation value of the
 lattice spin
\begin{equation}
  \label{eq:os1}
{\bm S}_{q,{\sf y}}({\sf x}) = \sqrt{2\alpha^3 |\lambda|} (-1)^{\sf x} ({\rm sign}(\lambda) \hat{\bm e}_{q} + (-1)^{\sf{y}}  \hat{\bm e}_{q+1}).
\end{equation}
This means that the two-point spin correlations reduce to
\begin{eqnarray}
  \label{eq:os2}
&&{\bm S}_{q,{\sf y}}({\sf x})\cdot {\bm S}_{q',{\sf y'}}({\sf x'}) = 2\alpha^3 |\lambda| (-1)^{\sf{x} + \sf{x'}} (\delta_{q',q} [1+(-1)^{\sf{y}+\sf{y'}}]  + \nonumber\\
&& + {\rm sign}(\lambda)[\delta_{q',q-1} (-1)^{\sf y'} + \delta_{q',q+1} (-1)^{\sf y}]).
\end{eqnarray}
An interesting feature of this result is that for $q=q'$, {\it i.e.} for different chains of the same kind,
${\bm S}_{q,{\sf y}}({\sf x})\cdot {\bm S}_{q,{\sf y'}}({\sf x'}) \sim [1+(-1)^{\sf{y}+\sf{y'}}]$.  That is, spins from the like chains
of opposite parity (when $\sf{y}+\sf{y'} = {\text{odd}}$) are orthogonal to each other. This peculiar feature of the cuboc order
is clearly seen in panels (a) and (b) of Fig.~\ref{fig:xcspin}.
Note also that while the same-chain spin correlations are not sensitive
to the sign of $\lambda$, those between the spins with different $q$'s are proportional to $\lambda$, and take opposite values
in cuboc1 and cuboc2 phases. This feature too is visible in the numerical data of Fig.~\ref{fig:xcspin}.

Spin chirality can be analyzed similarly. We find
\begin{eqnarray}
  \label{eq:os3}
&&{\bm S}_{1,{\sf y_1}}({\sf x_1})\cdot {\bm S}_{2,{\sf y_2}}({\sf x_2}) \times {\bm S}_{3,{\sf y_3}}({\sf x_3})= (2\alpha^3 |\lambda|)^{3/2} (-1)^{\sf{x_1} + \sf{x_2} +
\sf{x_3}}\nonumber\\
&&\times[{\rm sign}(\lambda) + (-1)^{\sf{y_1} + \sf{y_2} + \sf{y_3}}] \hat{\bm e}_{1}\cdot \hat{\bm e}_{2}\times \hat{\bm e}_{3}.
\end{eqnarray}
This shows that chiralities $\chi_{\triangle_1,\triangle_3,\triangle_4}$ acquire finite (and different) expectation values in the two cuboc phases, see Fig.~\ref{fig:xccor}.
At the same time within the chain mean-field  $\chi_{\triangle_2} = 0$ because it involves two spin from the same chain, which nullifies the triple product of spins identically.
Numerical data in Fig.~\ref{fig:xccor} does show somewhat suppressed but certainly not zero $\chi_{\triangle_2}$. This, we think, happens due to contributions from the
subleading
uniform part $\hat{\bm M}_{q,{\sf y}}({\sf x})$ of the spin operator, Eq.~\eqref{eq:1}, which is not captured by the mean-field treatment.

 The chain mean-field approach completely neglects {\em marginal} interchain interaction of spin currents,
 $4J_1 \sum_{\sf y,y'} {\bm M}_{q,{\sf y}}(-{\sf y}') {\bm M}_{q+1,{\sf y}'}({\sf y}+{\sf y}')$, see Eq.~\eqref{eq:5}. This approximation is truly justified only in the weak coupling $J_{1,2} \ll J_d$ limit, when the logarithmically slow grows of the marginal coupling constant (scaling dimension 2) is certain to {\em not} spoil
 an exponentially faster grows of the relevant ${\bm N}_{q} \cdot {\bm N}_{q+1}$ term (scaling dimension 1). However, a ferromagnetic (negative) sign of the
 interchain interactions $J_{1,2}$ is known to change this marginal interaction into a marginally irrelevant (logarithmically decaying) one, which has the effect of extending the range of validity of the chain mean-field approximation \cite{Hikihara2010}.

\subsubsection{Compensated regime}
\label{sec:compensated-regime}

Near the line $J_1=J_2$, the leading coupling $\lambda$ vanishes.
Here it is necessary to include the subdominant term in
Eq.~\eqref{eq:5}.  Owing to the derivative and the factor of
${\bm M}_{q,{\sf y}}$, it appears to be strongly irrelevant, and
na\"ively one might expect that the decoupled chain state is stable.
In reality, it generates a more subtle instability towards VBC order.  This
occurs by a fluctuation effect: the irrelevant coupling in
Eq.~\eqref{eq:5} generates a relevant one at second order upon
renormalization.

The procedure for calculating this fluctuation correction was worked
out in Ref.~\onlinecite{starykh2005anisotropic}.  We work with the imaginary
time path integral and expand the weight $e^{-S}$ to second order in
the interaction part of the action $S' =
\int\! d\tau\, H'|_{\rm sub-dom}$, and use the fusion rules of the
current algebra of SU(2)$_1$ to perform the renormalization.  The
correction to the effective action is
\begin{equation}
  \label{eq:14}
  \delta S = -\frac{1}{2} \int \! d\tau\, d\tau' \left[H'|_{\rm
    sub-dom}(\tau) H'|_{\rm sub-dom}(\tau')\right]_>,
\end{equation}
where the brackets $[\cdot]_>$ indicates renormalization by removing
of high energy/short-time degrees of freedom.  The dominant effect
comes from the cross-term,
\begin{widetext}
\begin{eqnarray}
  \label{eq:15}
  \delta S & \sim & -\frac{(J_1)^2}{2} \sum_{{\sf y},{\sf y}',q} \int \! d\tau\,
  d\tau' \Big[
\left(2{\bm M}_{q,{\sf y}}(-{\sf y}',\tau) +(-1)^{{\sf y}'} \partial_{\sf x}{\bm N}_{q,{\sf
         y}}(-{\sf y}',\tau)\right) \cdot \left(2 {\bm M}_{q+1,{\sf
         y}'}({\sf y}+{\sf y}',\tau) + (-1)^{{\sf y}+{\sf y}'}\partial_x{\bm N}_{q+1,{\sf y}'}({\sf
       y}+{\sf y}',\tau) \right)\nonumber \\
  && \times
\left(2{\bm M}_{q,{\sf y}}(-{\sf y}',\tau') +(-1)^{{\sf y}'} \partial_{\sf x}{\bm N}_{q,{\sf
         y}}(-{\sf y}',\tau')\right) \cdot \left(2 {\bm M}_{q+1,{\sf
         y}'}({\sf y}+{\sf y}',\tau') + (-1)^{{\sf y}+{\sf y}'}\partial_x{\bm N}_{q+1,{\sf y}'}({\sf
       y}+{\sf y}',\tau') \right)
\Big]_> \nonumber \\
  & = &   -\frac{(J_1)^2}{2} \sum_{a,b=x,y,z}\sum_{{\sf y},{\sf y}',q} \int \! d\tau\,
  d\tau' \Big[\left(2M^a_{q,{\sf y}}(-{\sf y}',\tau) +(-1)^{{\sf y}'} \partial_{\sf x}N^a_{q,{\sf
         y}}(-{\sf y}',\tau)\right) \left(2M^b_{q,{\sf y}}(-{\sf
        y}',\tau') +(-1)^{{\sf y}'} \partial_{\sf x}N^b_{q,{\sf
        y}}(-{\sf y}',\tau')\right) \Big]_> \nonumber \\
  && \times \Big[\left(2 M^a_{q+1,{\sf
         y}'}({\sf y}+{\sf y}',\tau) + (-1)^{{\sf y}+{\sf
     y}'}\partial_x N^a_{q+1,{\sf y}'}({\sf
       y}+{\sf y}',\tau) \right) \left(2 M^b_{q+1,{\sf
         y}'}({\sf y}+{\sf y}',\tau') + (-1)^{{\sf y}+{\sf
     y}'}\partial_x N^b_{q+1,{\sf y}'}({\sf
       y}+{\sf y}',\tau') \right)\Big]_>
\end{eqnarray}
We use the identity, from Eq.~(35) of
Ref.~\onlinecite{starykh2005anisotropic},
\begin{equation}
  \label{eq:16}
  \left[ M^a({\sf x},\tau) \partial_x N^b({\sf x},\tau')\right]_> = -
  \frac{\delta^{ab} \varepsilon({\sf x},\tau)}{2\pi v^2
    (\tau-\tau'+\tau_0 \sigma_{\tau-\tau'})^2},
\end{equation}
where $v= \pi J_d/2$ is the velocity of the 1d Heisenberg chain, $\tau_0
\sim 1/J_d$ is a short-time cutoff, and $\sigma_\tau = {\rm
  sign}(\tau)$.  Using this gives
\begin{eqnarray}
  \label{eq:17}
   \delta S & \sim & -8 J_1^2\sum_{a,b}\, \sum_{{\sf y},{\sf y}',q} \int \!
                     d\tau\,   d\tau'
(-1)^{{\sf y}} \frac{\delta^{ab} \varepsilon_{q,{\sf y}}(-{\sf y}',\tau)}{2\pi v^2
    (\tau-\tau'+\tau_0 \sigma_{\tau-\tau'})^2}
                     \frac{\delta^{ab} \varepsilon_{q+1,{\sf y}'}({\sf y}+{\sf y}',\tau)}{2\pi v^2
    (\tau-\tau'+\tau_0 \sigma_{\tau-\tau'})^2} \nonumber \\
  & = & -\frac{24 J_1^2}{(2\pi)^2} \sum_{{\sf y},{\sf y}',q} \int \!
                     d\tau\,  \left[ \int\! d\tau' \frac{1}{(v\tau' +
        \tau_0 v \sigma_{\tau'})^4}\right] (-1)^{{\sf y}}
        \varepsilon_{q,{\sf y}}(-{\sf y}',\tau) \varepsilon_{q+1,{\sf
        y}'}({\sf y}+{\sf y}',\tau) \nonumber \\
  & = & -\frac{4J_1^2}{\pi^2 v a_0^3} \sum_{{\sf y},{\sf y}',q} \int \!
                     d\tau\,   (-1)^{{\sf y}}
        \varepsilon_{q,{\sf y}}(-{\sf y}',\tau) \varepsilon_{q+1,{\sf
        y}'}({\sf y}+{\sf y}',\tau),
\end{eqnarray}
\end{widetext}
 where $\tau_0 v = a_0$ is a short-distance cut-off.  This can be
interpreted as the integral of a correction to the Hamiltonian,
\begin{equation}
  \label{eq:18}
  \delta H_{\rm int} = - \frac{4J_1^2}{\pi^2 v a_0^3} \sum_{{\sf
      y},{\sf y}',q}  (-1)^{{\sf y}}
        \varepsilon_{q,{\sf y}}(-{\sf y}') \varepsilon_{q+1,{\sf
        y}'}({\sf y}+{\sf y}').
\end{equation}
Note the distinct similarity to Eq.~\eqref{eq:3}, with staggered
dimerization $\varepsilon$ replacing the N\'eel operator ${\bm N}$.

The interaction $\delta H_{\rm int}$ should be added to that in
$H'_{\rm dom}$.  For $J_1=J_2$, it becomes the only important
interaction.    We analyze it using chain mean field theory, as we did
for the dominant interaction away from this line above.  We write the
variational Hamiltonian
\begin{equation}
  \label{eq:19}
  \tilde{H}_{\rm var} = H_0 - \sum_{q,{\sf y}} \int \! d{\sf x}\,
  \varphi_{q,{\sf y}} \varepsilon_{q,{\sf y}}({\sf x}).
\end{equation}
We have $\bar\varepsilon_{q,{\sf y}} = \langle
\varepsilon_{q,{\sf y}}\rangle = \tilde\alpha |\varphi_{q,{\sf
    y}}|^{1/3} {\rm sign} (\varphi_{q,{\sf y}})$, with
$\tilde\alpha>0$ another $O(1)$ constant, and the variational
energy is
\begin{equation}
  \label{eq:20}
  \tilde{E}_{\rm var} = \frac{\sf L}{4\tilde\alpha^3} \sum_{q,{\sf y}}
  \bar\varepsilon_{q,{\sf y}}^4 - \tilde\lambda \sum_q \sum_{{\sf y}, {\sf
      y}'} (-1)^{\sf y} \bar\varepsilon_{q,{\sf y}} \bar\varepsilon_{q+1,{\sf
        y}'}.
\end{equation}
Here we defined $\tilde\lambda = 4J_1^2/(\pi^2 v a_0^3)$.  Now define
\begin{equation}
  \label{eq:21}
   {V}_{q} = \frac{1}{\sl L} \sum_y (-1)^y \bar\varepsilon_{q,{\sf y}},
  \qquad {W}_q = \frac{1}{\sl L} \sum_y \bar\varepsilon_{q,{\sf y}}.
\end{equation}
Then the energy becomes
\begin{equation}
  \label{eq:22}
   \tilde{E}_{\rm var}/\mathcal{N} = \frac{1}{24\tilde\alpha^3} \sum_q
   \left[ (V_q+W_q)^4 + (V_q-W_q)^4\right] - \frac{\tilde\lambda}{3}
   \sum_q V_q W_{q+1}.
\end{equation}

\begin{figure}[h!]
  \centering
  \includegraphics[width=0.8\columnwidth, angle =90]{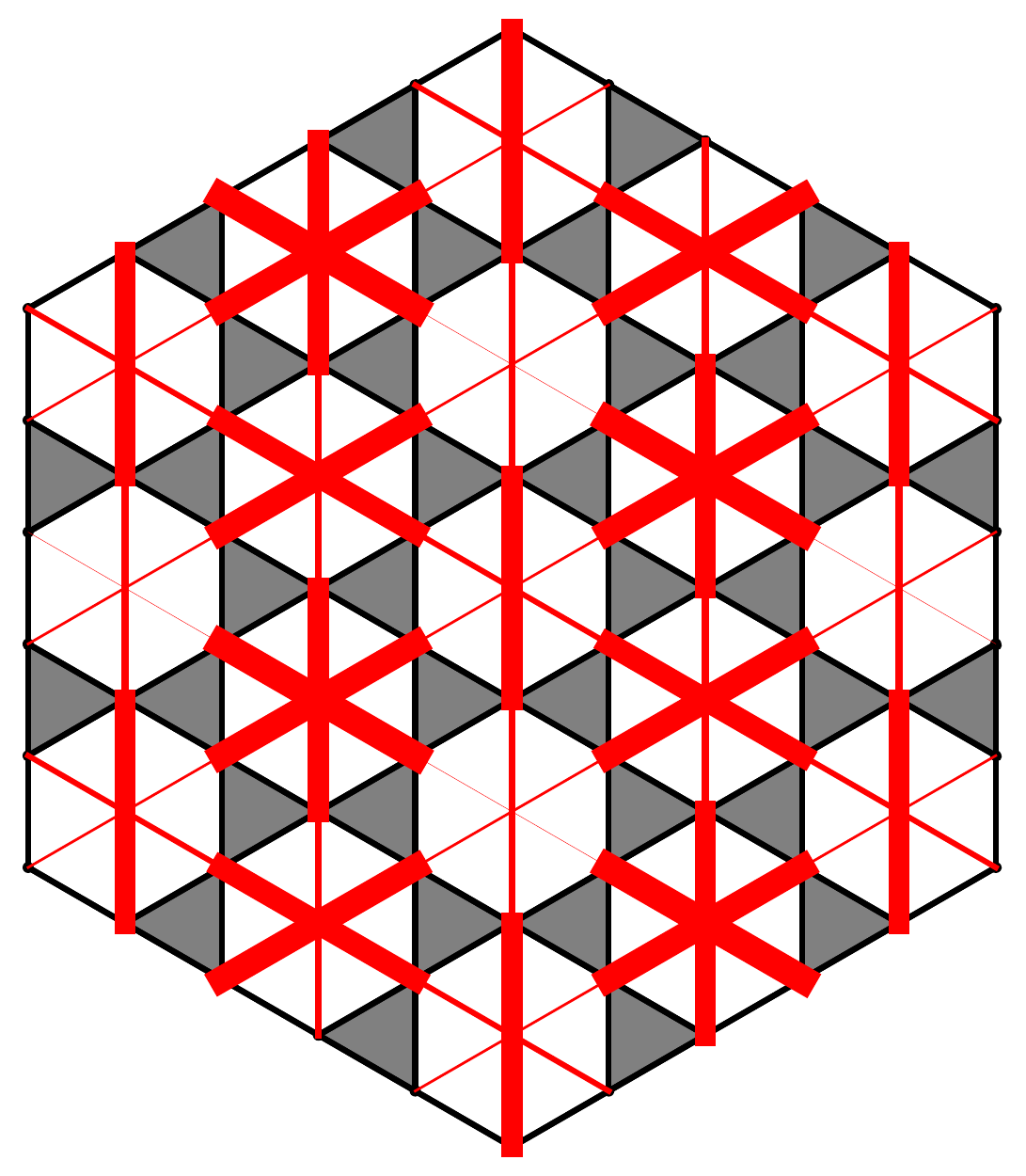}
  \caption{(Color online) Analytically predicted VBC pattern in two
    dimensions.  The thickness of each red line is proportional to the
    magnitude of the exchange on the bond. Observe that the unit cell
    is quadrupled.  This pattern may be compared to the one obtained
    by the DMRG calculation in Fig.~\ref{fig:dmrgvbc}(a). }
  \label{fig:vbc}
\end{figure}

All these manipulations and formulae look very similar to those we
carried out for magnetic ordering, but the minima of the variational
energy are quite different.  This is because $\bar\epsilon$ is a
scalar, and as a consequence the ordering is more frustrated.
Some algebra shows that the global minima of Eq.~\eqref{eq:22} are of
the form
\begin{eqnarray}
  \label{eq:23}
  (V_1,V_2,V_3)& =& \sqrt{ \tilde\alpha^3 \tilde\lambda}(0,s \upsilon,s'(
                    \upsilon')^3), \nonumber \\
  (W_1,W_2,W_3)& =&\sqrt{ \tilde\alpha^3 \tilde\lambda}
  (s'\upsilon',0, s (\upsilon)^3),
\end{eqnarray}
where $\upsilon = (7+3\sqrt{5})^{1/8}/\sqrt{2}\approx 0.981$ and
$\upsilon'=  (7-3\sqrt{5})^{1/8}/\sqrt{2}\approx 0.606$, or vice-versa, and $s$ and
$s'$ are independently and freely taken to be $\pm 1$, and finally,
the solution in Eq.~\eqref{eq:23} can also be cyclically permuted,
$V_q \rightarrow V_{q+1}, W_q \rightarrow W_{q+1}$.
This gives a set of 24 minima with equivalent energy,
$\tilde{E}_{\rm var} \approx -0.18 {\tilde \alpha}^3 {\tilde \lambda}^2$.  The
corresponding order is illustrated in Fig.~\ref{fig:vbc}.  By working out
the action of the space group symmetry of the kagom\'e lattice (see
Appendix~\ref{sec:symm-transf}),  we can check
that all these 24 states are related by symmetry.  So this degeneracy is
not accidental and is mandated.

\subsubsection{Phase boundaries}
\label{sec:phase-boundary}

Based on the above discussion, we expect the VBC phase on the line
$J_1=J_2$, and the two cuboc phases away from this line.  Since the VBC
state is a {\em phase}, it must exist in a finite width region around
the compensated line.  The shape of the boundaries
of this region is determined by comparing the VBC and magnetic
coupling constants, because the interactions in Eq.~\eqref{eq:3} and
Eq.~\eqref{eq:18} have the same scaling dimension.  This means the
boundaries occur when comparing Eqs.~\eqref{eq:13} and
\eqref{eq:23}, $\sqrt{|\lambda|} \sim \sqrt{\tilde\lambda}$.
Equivalently, we can compare the energies of the two orders, $E_{\rm var} = \tilde{E}_{\rm var}$.
Approximating the ratio
$\alpha/\tilde\alpha \sim 1$ and using $v a_0 = \pi J_d/2$, we obtain
\begin{equation}
  \label{eq:32}
  \frac{J_1}{J_d} - \frac{J_2}{J_d} = \pm 0.055 \Big(\frac{J_1}{J_d}\Big)^2 .
\end{equation}
This is only a scaling relation -- the numerical prefactor above is an
estimate based on a weak-coupling analysis.  Its smallness would
imply a region of VBC phase narrower than the DMRG numerics of
Sec.~\ref{sec:dmrg-results} would indicate, suggesting the true
prefactor may be larger.
However, Eq.~\eqref{eq:32} is enough to show that the VBC phase occupies a
wedge-shaped region around the diagonal in the $J_1-J_2$ plane, which
pinches down to zero width as the origin is approached, as is sketched in the
phase diagram in Fig.~\ref{fig:phase}.

\subsection{Cylinders}
\label{sec:cylinders}

To compare with DMRG calculations on finite circumference ($L_y$) cylinders,
we consider this type of geometry explicitly.  In general, a
complicated dependence may exist on the choice of embedding the
lattice into the cylinder, and also the cylinder circumference.  For
small cylinders, the finite size effects can be quite substantial.
This is especially expected when there is a long length scale already
in the two dimensional problem, for example in the limit
$J_1,J_2 \ll J_d$.  Then we need to compare the circumference to this
two dimensional correlation length.

\subsubsection{Quasi-2d limit}
\label{sec:quasi-2d-limit}

In the limit of very large circumference cylinders, i.e. when the
circumference is large compared to all two dimensional correlation
lengths, then we expect relatively universal behavior.  In this case,
the system is essentially ordered on scales smaller than the cylinder
width, and the only important degrees of freedom on those scales and
larger are captured by the order parameter(s).  We will call this the
quasi-2d limit.

If the two dimensional system is in one of the magnetically ordered
cuboc states, then there are two order parameters.  One is the
discrete chiral order parameter, which is of Ising type.  The other is
the continuous SO(3) order parameter that specifies the specific spin
orientations.  The former chiral order parameter has only gapped
fluctuations, and would be expected can retain its order at $T=0$ in
the limit of wide cylinders, even as the length of these cylinders
extends to infinity.  So {\em we expect spontaneous chiral order in
  sufficiently wide cylinders away from the compensated line}.  The
SO(3) order parameter, being continuous, by contrast cannot
spontaneously order in one dimension.  It instead is governed by an
SO(3) matrix non-linear sigma model in 1+1 dimensions with a small
coupling constant $g \sim 1/L_y$ (effective ``temperature'' for 2d
Euclidean theory), which is expected to be asymptotically free.  Some
gapped behavior, with exponentially decaying spin correlations beyond
some {\em one-dimensional} correlation length
$\xi_{1d} \sim \exp(g_0/g)$ should be expected.
This may be accompanied
by spontaneous dimerization, whose presence may depend upon the parity
of the circumference.

This argument also suggests that for sufficiently small $L_y$
(that is, high effective temperature) the chiral order will melt as well,
leading to a state without long range order in both spin and chiral degrees of freedom.
This is perhaps what is observed in cuboc regions in both XC and YC geometries (see
the definitions in Fig.~\ref{fig:lattice}), see
Figs.~\ref{fig:xccor} and \ref{fig:yccor}, panels (c) and (d).

In the vicinity of the compensated line, we expect VBC order in the
infinite 2d limit.  Since the VBC phase has an entirely discrete order
parameter, we expect symmetry breaking to remain for finite width
cylinders.  The analysis is involved, however, since the
symmetries broken by the VBC order are space group operations, some of
which are broken by confinement to the cylinder, in ways which depend
upon the geometry and circumference of the cylinder. A further complication is that,
for some cylinders, notably of odd circumference, the 2d VBC order may
be incompatible with periodic boundary conditions around the
cylinder.  In this case, defects such as domain walls may be present,
and the gap may close at these defects.  We will eschew any detailed
analysis beyond these general remarks.

\subsubsection{Quasi-1d Limit: YC cylinders}
\label{sec:quasi-1d-limit}

If the cylinder circumference is not too large, it can interfere with
even the short-range development of order.  The effect is particularly
clear for the YC cylinders, in which one of the three types of chains
-- we choose this to be type ``1'' for concreteness -- is oriented
along the periodic direction.  The type 1 chains are therefore
finite in the YC geometry.  For such finite chains, even without any
inter-chain coupling, the spins form a gapped singlet state, with a
gap of order $v/L_y$.  We can expect that if the inter-chain coupling
is in some sense weaker than this finite size gap, the spins on these
chains will resist ordering.

\begin{figure}[h!]
  \centering
  \includegraphics[width=0.8\columnwidth]{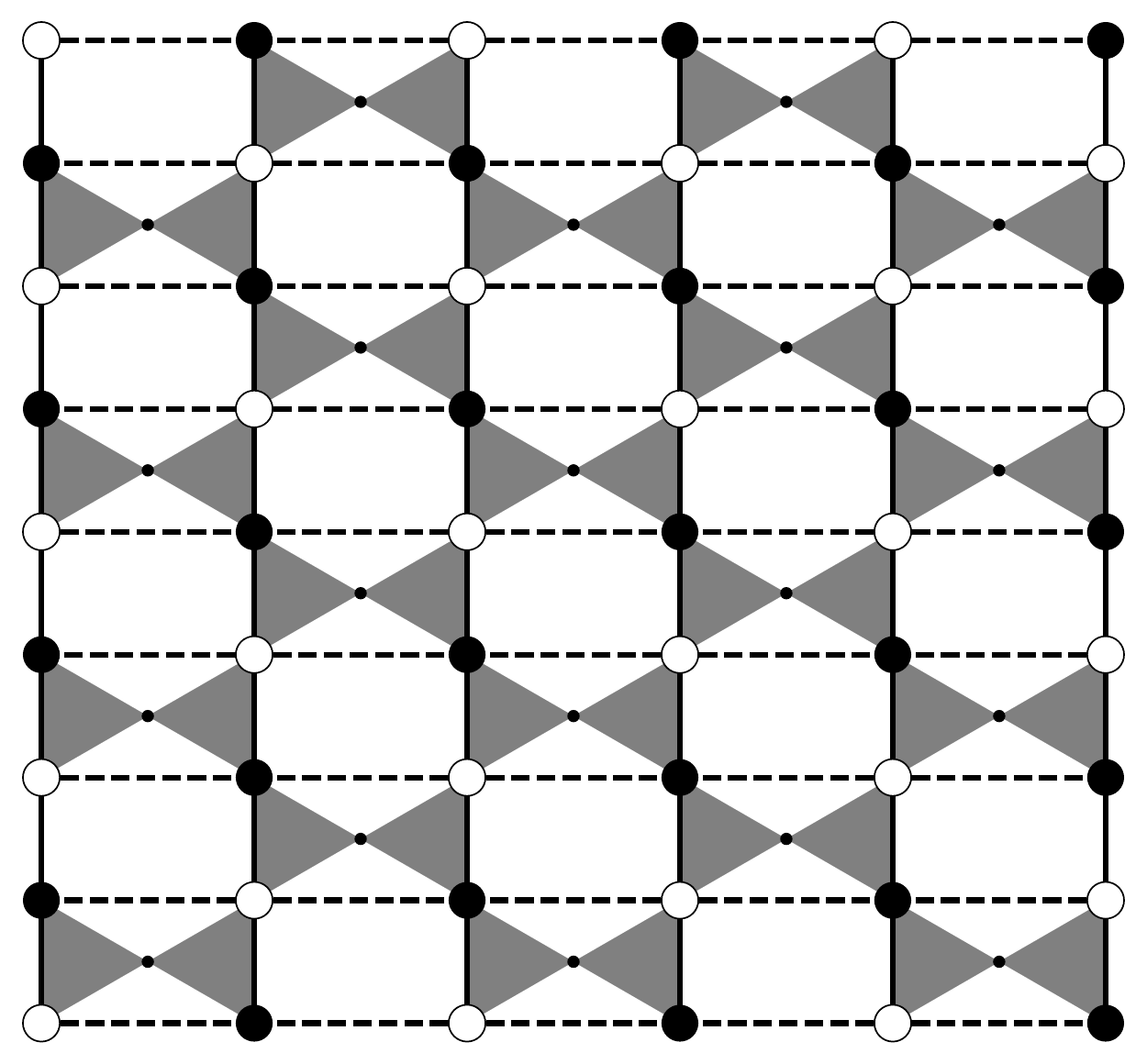}
  \caption{Two types of gapless chains in the YC cylinder
    geometry, and their couplings.  White and black circles represent
    the sites of two types of infinite chains.  Solid vertical bonds
    and dashed horizontal ones are the $J_1$ and $J_2$ couplings
    between these sites.}
  \label{fig:YCtwosub}
\end{figure}

At a first level of analysis, we can understand the physics of this
limit by simply neglecting the type 1 spins.  Dropping the ${\bm
  N}_{q,{\sf y}}$ terms in Eq.~\eqref{eq:3}, we obtain unfrustrated
interactions between the N\'eel fields on the remaining chains type
$2$ and $3$ chains, which
actually favors {\em collinear} ordering.  This is also apparent from
a visual inspection of the geometry of the 1-2 sublattices alone (see
Fig.~\ref{fig:YCtwosub}).  One observes that a ferromagnetic $J_1$
interaction is unfrustrated and results in ferromagnetic alignment of
spins in each vertical column, with successive columns aligned
antiferromagnetically, due to the strong antiferromagnetic $J_d$
coupling.  In this pattern all $J_1$ and $J_d$ interactions are
perfectly satisfied.  A ferromagnetic $J_2$ induces instead ferromagnetic
alignment of each horizontal row of spins, with antiferromagnetic
alignment of successive rows by $J_d$.  Here all $J_2$ and $J_d$
interactions are satisfied.

The conclusion is that for the YC cylinders of ``small''
circumference, there is a strong finite size effect which favors {\em
  collinear} order rather than the non-collinear cuboc type.  Spins on
the type 1 chains are gapped by finite size effects, and exhibit
exponential decay of correlations along the cylinder with a
correlation length of order a single lattice site.   In either
collinear pattern the net exchange field on the type 1 sites vanishes,
so there is no induced moment there.  The collinearity and absence of
a moment on the type 1 site means that the chirality is suppressed,
and should exhibit exponential correlations along the cylinder,
similar to that of the type 1 spins, and even further suppressed by
the collinearity of the type 2+3 spins.

For a very long cylinder of finite circumference, the chain mean field
theory must be further corrected for one-dimensional quantum
fluctuations.  This of course prohibits any type of, including
collinear, N\'eel long range order.  Instead the system will be
governed by a vector SO(3) non-linear sigma model, and we expect only
a trivial $\Theta$ term for the even circumference cylinders we study
here.  Ultimately this will induce a small gap and exponential decay
of correlations also on the type 2 and 3 sites, but with a much longer
correlation length.  However, on observable short distances we expect
to observe behavior quite compatible with two dimensional N\'eel
orders of the types indicated above. The data in
the ``magnetically ordered'' regime from the DMRG on YC cylinders
fits very well to such collinear behavior for small
circumference, as Fig.~\ref{fig:ycspin} shows.

\begin{figure}[h!]
  \centering
    \includegraphics[width=0.8\columnwidth]{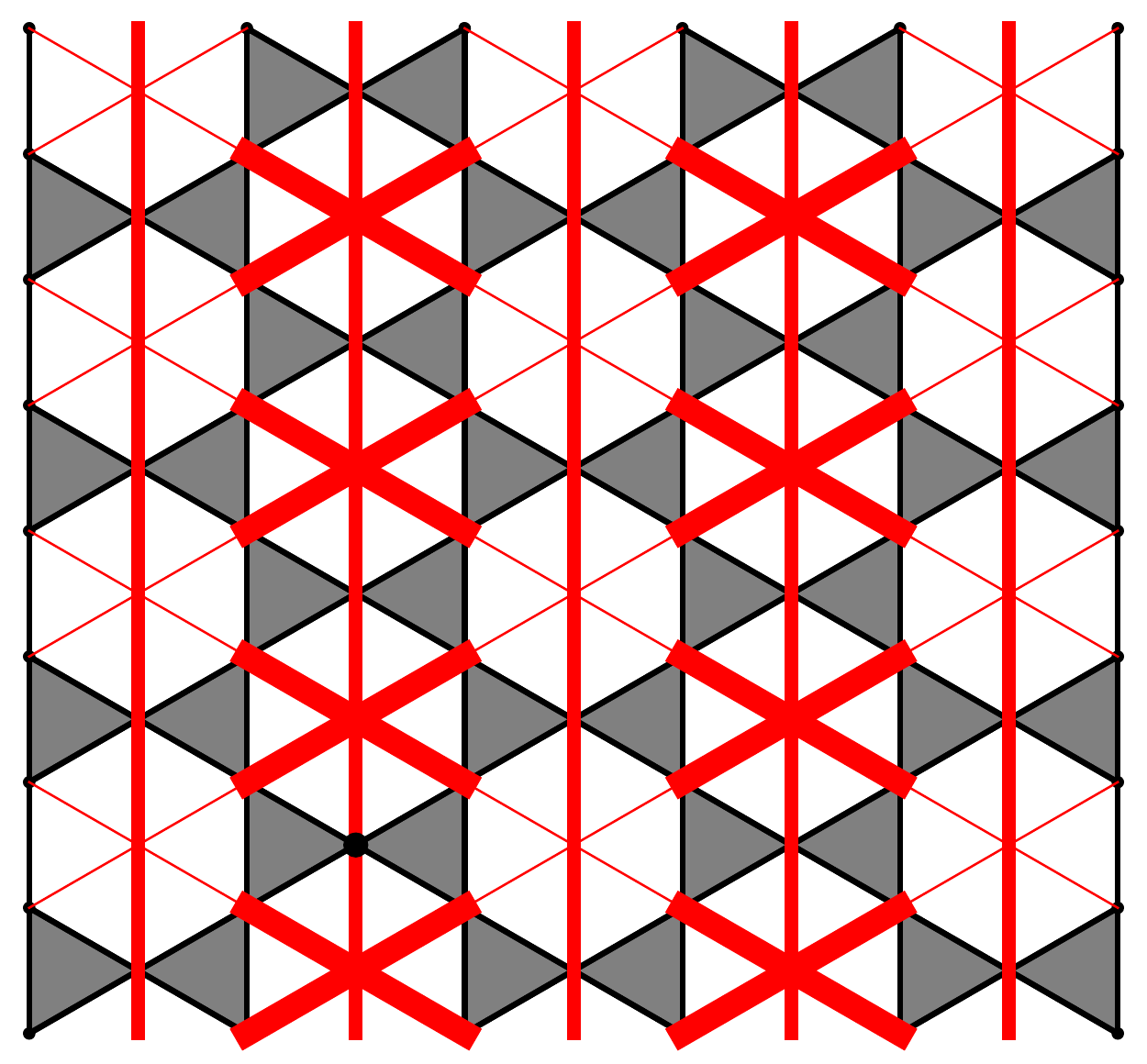}
  \caption{The analytically predicted VBC state on YC cylinders, which
  differs from that in Fig.~\ref{fig:vbc} due to strong finite size
  effects.  This should be compared to the DMRG result in Fig.~\ref{fig:dmrgvbc}(b).}
  \label{fig:ycvbc}
\end{figure}

{\em VBC state in the YC cylinder:} On the compensated line $J_1=J_2$, we again
need to consider the fluctuation-induced dimerization interactions of
Eq.~\eqref{eq:18}, but now examine its effects on the finite
cylinder.  Due to the smaller magnitude of the induced dimerization
coupling, the finite size effects on the short type 1 chains is even
more significant -- the finite size gap is larger relative to this
interaction.  Hence in this geometry, we should simply neglect
dimerization of the type one chains, and set $\bar\varepsilon_{1,{\sf
y}}=0$.  Then the minimum of the chain mean field variational
energy is simply
\begin{equation}
  \label{eq:35}
  \bm{V} = \sqrt{{\tilde \alpha}^3 \tilde\lambda} (0,s,0), \qquad \bm{W} =  \sqrt{{\tilde \alpha}^3 \tilde\lambda} (0,0,s),
\end{equation}
with $s=\pm 1$ defining two degenerate solutions.  The resulting VBC
pattern, shown in Fig.~\ref{fig:ycvbc}, is translationally
invariant in the vertical ($1$) direction along the cylinder
circumference, and has period two normal to it.  The two signs of the
solution simply represent these translational copies. Note that the
strong finite size effects have greatly simplified the VBC order
relative to the 24-fold degenerate state expected in two dimensions.
The DMRG results for YC cylinders at $J_1=J_2$ seems most consistent with
this simpler VBC order, see Fig.~\ref{fig:dmrgvbc}(b).

The presented arguments make us conclude that XC cylinders, results
on which are described in details in the next section, approximate the desired 2d limit
of the model better than those of YC kind. Nonetheless the seemingly exponential
decay of the chirality correlations in Fig.~\ref{fig:xccor}, panels (c) and (d), suggests
that the studied XC cylinders are still too narrow to truly capture the 2d physics of the
non-coplanar cuboc phase.

\section{DMRG results}
\label{sec:dmrg-results}

\begin{figure}[h!]
  \includegraphics[width=1.0\columnwidth]{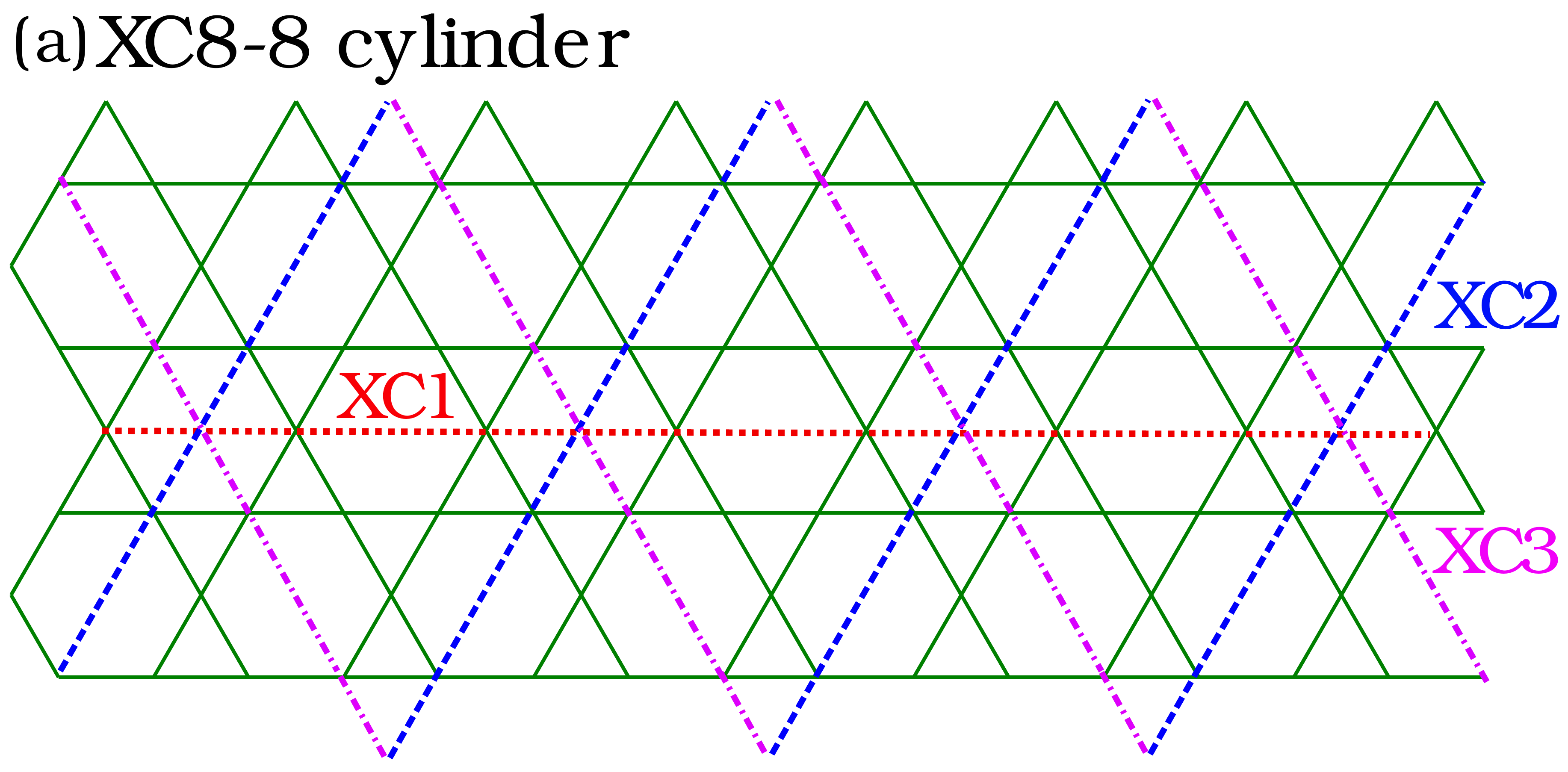}
  \includegraphics[width=1.0\columnwidth]{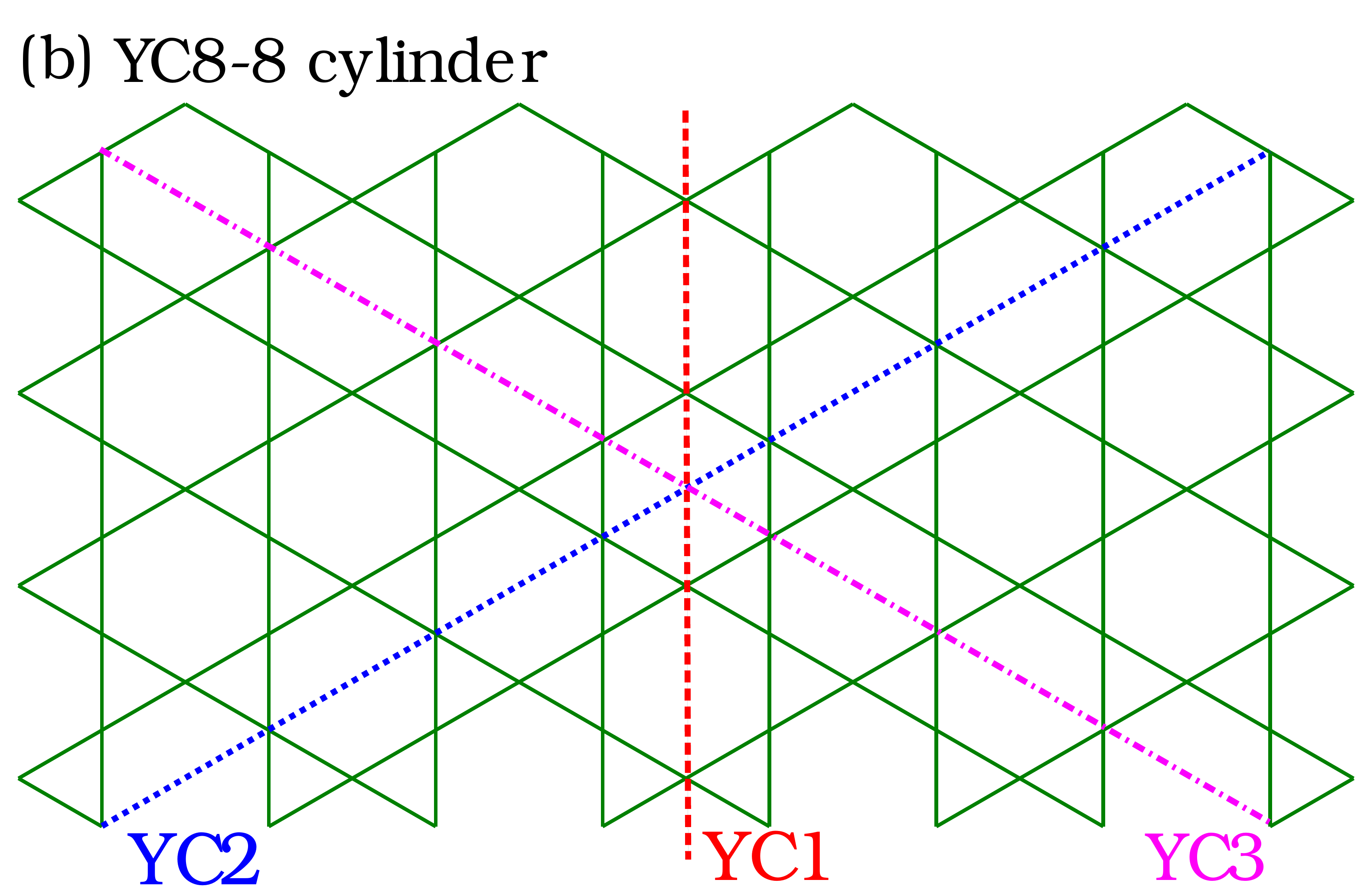}
  \caption{(Color online) Cylinder geometries used in the DMRG
    calculations.  (a) A XC8-8 cylinder on kagom\'{e} lattice.  This
    cylinder has $4$ unit cells along the $y$ direction and $8$ unit
    cells along the $x$ direction.  The three dashed lines denote the
    $J_d$ chains along the three directions.  This cylinder has $4$
    XC1 chains with $8$ sites along the $x$ direction and $2$ XC2 and
    XC3 chains with $16$ sites along the tilted directions.  All the
    $J_d$ chains are extended, i.e. have a length proportional to the
    long dimension of the cylinder.  (b) A YC8-8 cylinder on
    kagom\'{e} lattice.  This cylinder has $4$ unit cells along the
    $y$ direction and $8$ unit cells along the $x$ direction.  The
    three dashed lines denote the $J_d$ chains along the three
    directions.  This cylinder has $9$ YC1 chains with $4$ sites along
    the $y$ direction and $4$ YC2 and YC3 chains with $8$ sites along
    the tilted directions.  While the YC2 and YC3 chains are extended,
    the YC1 chains are closed with a short circumference along the $y$
    direction.}
  \label{fig:lattice}
\end{figure}

Here we report results of the numerical density-matrix
renormalization group \cite{white1992} (DMRG) studies of the
$J_1-J_2-J_d$ kagom\'{e} model.  Through calculations on cylinders, we
establish the quantum phase diagram as shown in Fig.~\ref{fig:phase},
which has two cuboctohedral phases, cuboc1 and cuboc2, separated by a
VBC phase region. We use a DMRG algorithm with spin rotational $SU(2)$
symmetry \cite{mcculloch2002} by keeping a number of $U(1)$-equivalent
states as large as $32000$.  To mitigate and understand finite size
effects, we study two different cylinder geometries denoted as XC and
YC, which have one of the three bond orientations along the $x$ and
$y$ axes, respectively (see Fig.~\ref{fig:lattice}).  The system size
is denoted as XC$2L_y$-$L_x$ and YC$2L_y$-$L_x$, where $L_y$ ($L_x$)
is the number of unit cells in the $y$ ($x$) direction.  We study the
YC cylinders with $L_y=4,6$ (YC8 and YC12) and XC cylinders with
$L_y=4$ (XC8). We do not study the XC12 cylinder ($L_y=6$), because this
geometry does not accommodate the cuboc ordering pattern. For the XC8 and YC8
cylinders, we obtain the converged energy with DMRG truncation error
$\sim 1\times 10^{-6}$ by keeping about $16000$ $U(1)$-equivalent states. 
For the YC12 cylinder, the truncation error can
only be reduced to about $5\times 10^{-5}$. Although the calculations
are not well converged for the YC12 cylinder, the results for this system are
qualitatively consistent with those for the YC8 cylinder.

\begin{figure}[h!]
  \includegraphics[width=1.0\columnwidth]{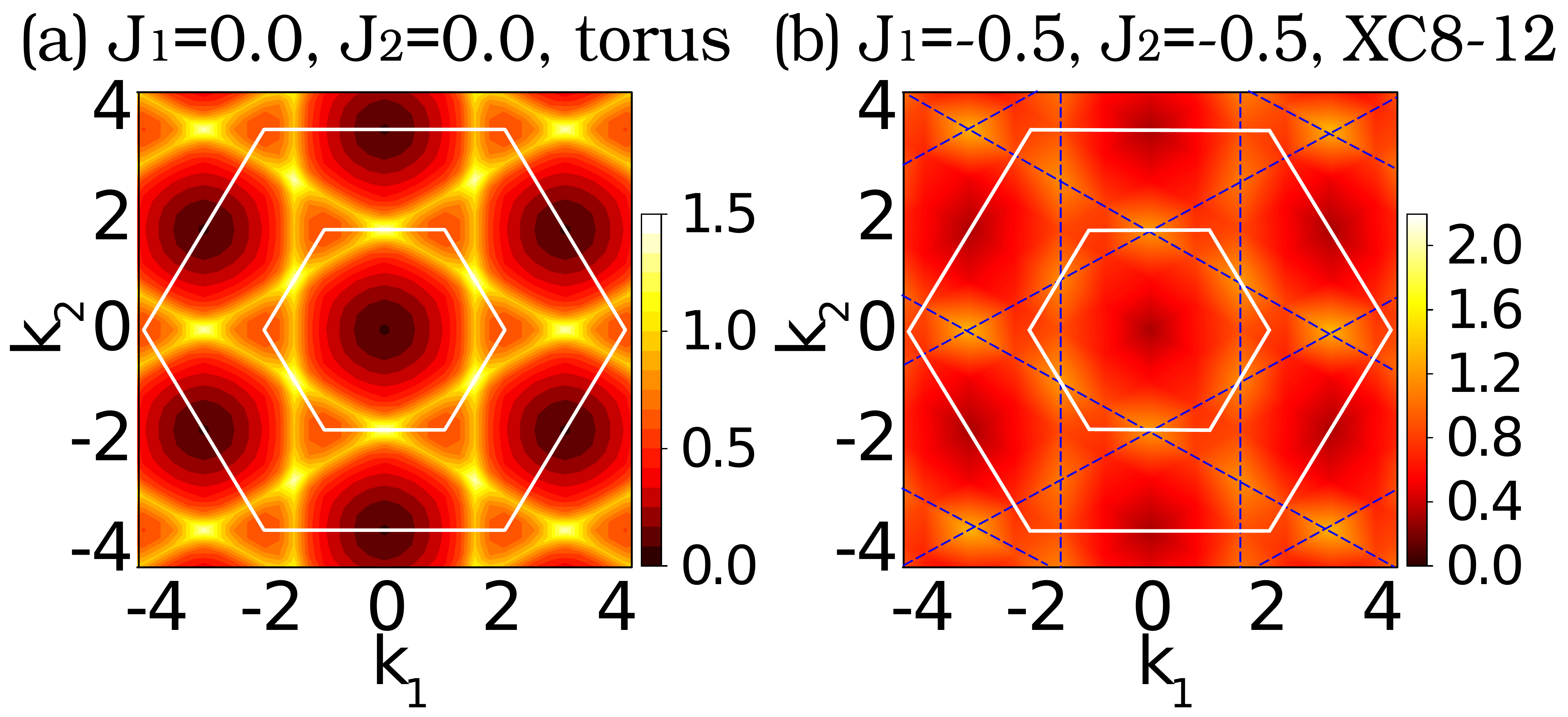}
  \includegraphics[width=1.0\columnwidth]{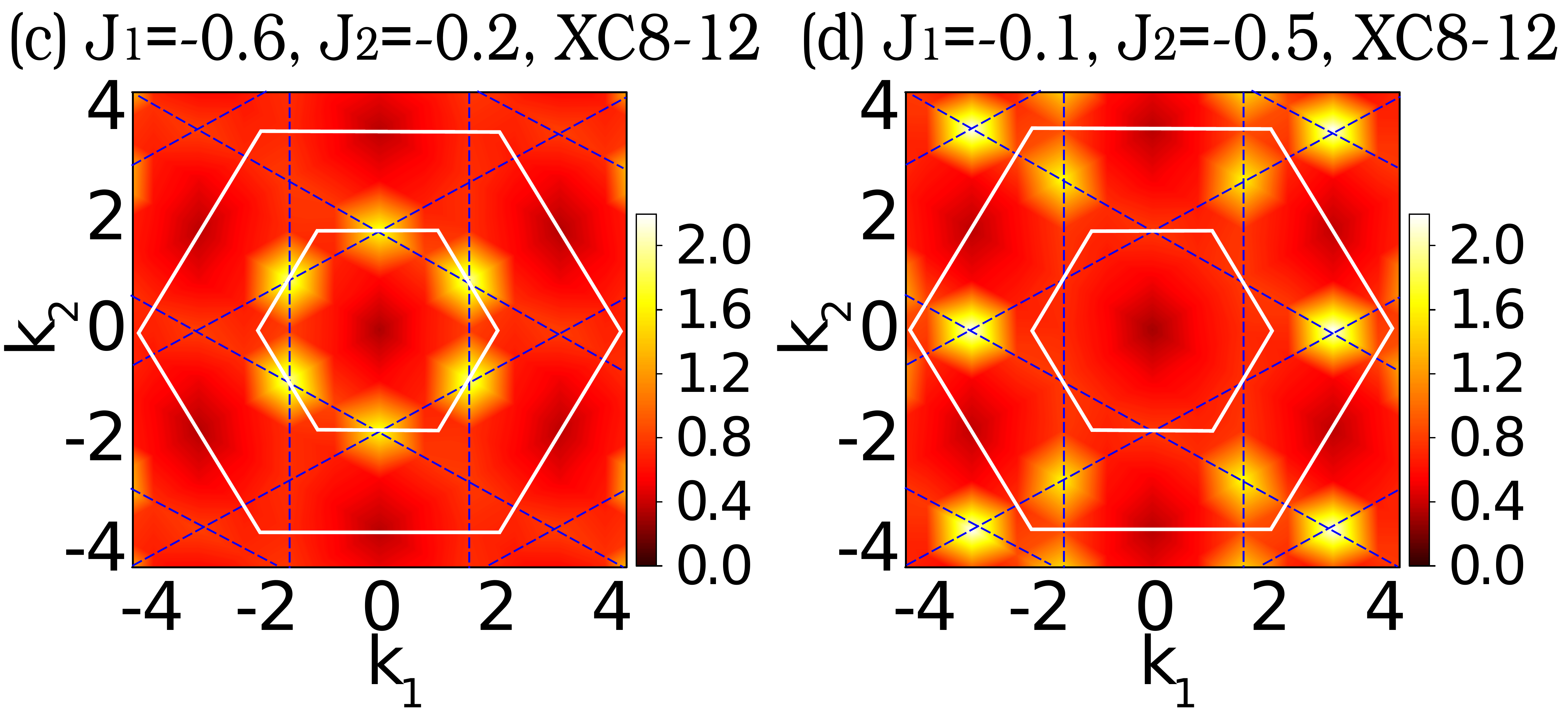}
  \caption{(Color online) Magnetic structure factor $S(\mathbf{k})$ in
    the extended Brillouin zone of the extended triangular
    lattice. The lattice spacing in real space is the length of the
    $J_1$ bond.  (a) $S(\mathbf{k})$ of the decoupled $J_d$ chains
    ($J_1=J_2=0.0$) on the $N = 3 \times 12\times 12$ torus.  (b)-(d)
    $S(\mathbf{k})$ obtained on XC8-12 cylinder for different
    phases. The blue dashed lines denote the momenta of the decoupled
    $J_d$ chain system.  In the compensated regime (b),
    $S(\mathbf{k})$ has the peaks at the same momenta as the decoupled
    $J_d$ chains as shown in (a), but is much smoother.  In the cuboc2
    phase (c), the peak intensity of the structure factor is at the
    six inner crossings, while in the cuboc1 phase (d), it is at the
    outer line crossings.  Due to the anisotropy inherent to the
    cylinder geometry, the magnitude of the six peaks of
    $S(\mathbf{k})$ are different.}
  \label{fig:sq}
\end{figure}

To obtain a coarse understanding of the phase diagram, we calculate
and compare the magnetic structure factor
$S(\mathbf{k}) = \frac{1}{N} \sum_{i,j} e^{i\mathbf{k}\cdot
  (\mathbf{r}_i - \mathbf{r}_j)}\langle \mathbf{S}_i \cdot
\mathbf{S}_j\rangle$
in different parameter regions for the XC cylinders.  We define the
Brillouin zone based on an extended triangular lattice in real space
\cite{fak2012}, which has a virtual site in the center of each hexagon
of the kagom\'{e} lattice. We define the lattice spacing in real space
as the nearest $J_1$ bond length.  Thus in the plots in
Fig.~\ref{fig:sq}, which show representative structure factors, the
smaller white hexagon is the Brillouin zone of the kagom\'{e} lattice
and the larger one is the Brillouin zone of the extended triangular
lattice. The dashed blue lines denote the momentum space of the
decoupled one-dimensional $J_d$ chains.  When calculating the
structure factor, the spin correlations including the virtual sites
are all set to zero.

Fig.~\ref{fig:sq}(a) shows the structure factor for $J_1 = J_2 = 0$,
in which the system consists of decoupled chains.  In this case it
exhibits peaks at the momenta with
$\mathbf{k} \cdot \mathbf{a}_q = \pi$ (the $\mathbf{a}_q$ with
$q=1,2,3$ were defined in Sec.~\ref{sec:mapp-coupl-chains}). One can
observe that the peak momenta are the crossing points of the momentum
lines.  The remaining three plots, Figs.~\ref{fig:sq}(b,c,d) show the
structure factor with non-zero interchain coupling.  Notably, in all
cases, the peaks of the structure factor coincide some subset of those
of decoupled chains, which indicates the appropriateness of the
treatment of the system in Sec.~\ref{sec:analytical-treatment}.  In
the compensated region with $J_1 \simeq J_2$, Fig.~\ref{fig:sq}(b),
all the features of the decouple chain structure factor are preserved
-- both high intensities along lines and peaks at their intersections.
However, the features themselves are broadened, and the structure
factor appears much less singular.  This suggests the system remains a
liquid as in the decoupled chains case, but with short-range rather
than power law spin correlations.  In the parameter region far from
the compensated line $J_1 = J_2$, we find that, as shown in
Figs.~\ref{fig:sq}(c,d), the peaks of $S(\mathbf{k})$ locate at the six
inner crossings or at the outer crossings depending for
$|J_1| > |J_2|$ or $|J_1| < |J_2|$, respectively.  The selection of
the inner and outer points agrees with the cuboc phases as shown in
the insets of Fig.~\ref{fig:cuboc}.  One further observes in
Figs.~\ref{fig:sq}(c,d) that two of the six peaks have larger
intensity, which must be attributed to the rotational symmetry
breaking induced by the cylindrical geometry.  This physics is
discussed in Sec.~\ref{sec:quasi-1d-limit}, and we return to it in
Sec.~\ref{sec:yc-cylinders}.  We now discuss the
DMRG results in detail for each phase region.

\subsection{Magnetically ordered region}
\label{sec:magnetic-orders}

First of all, we study the phase regions with the cuboc-like magnetic structure factors,
which are far from the compensated line $J_1 = J_2$.

\subsubsection{XC cylinders}

\begin{figure}[h!]
  \includegraphics[width=1.0\columnwidth]{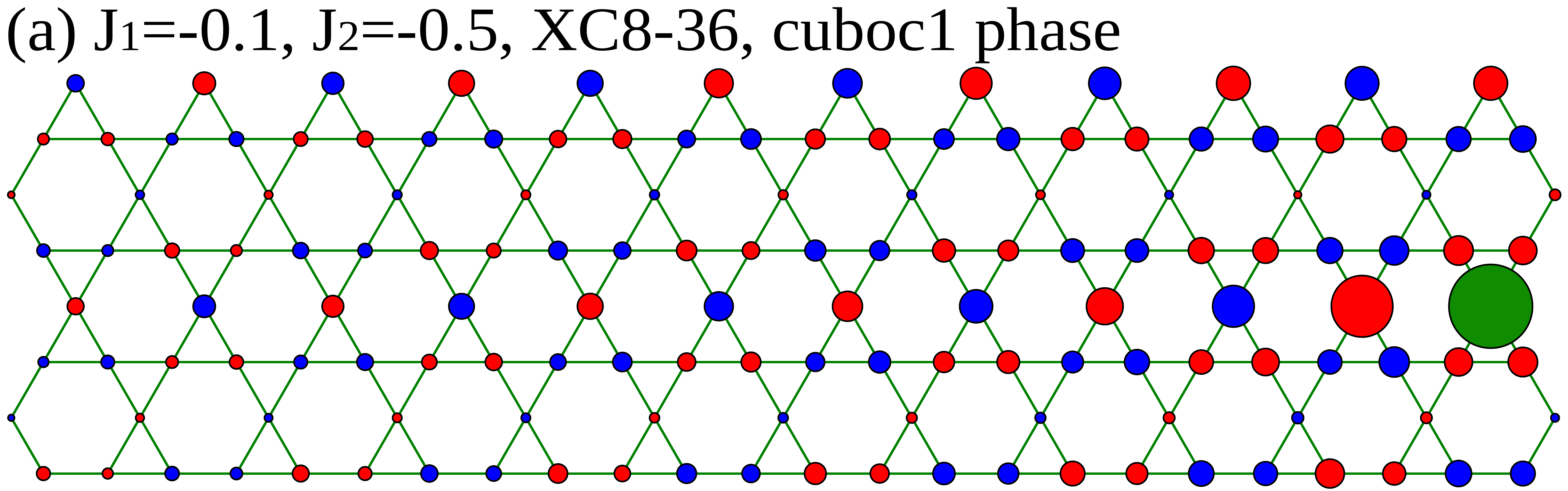}
  \includegraphics[width=1.0\columnwidth]{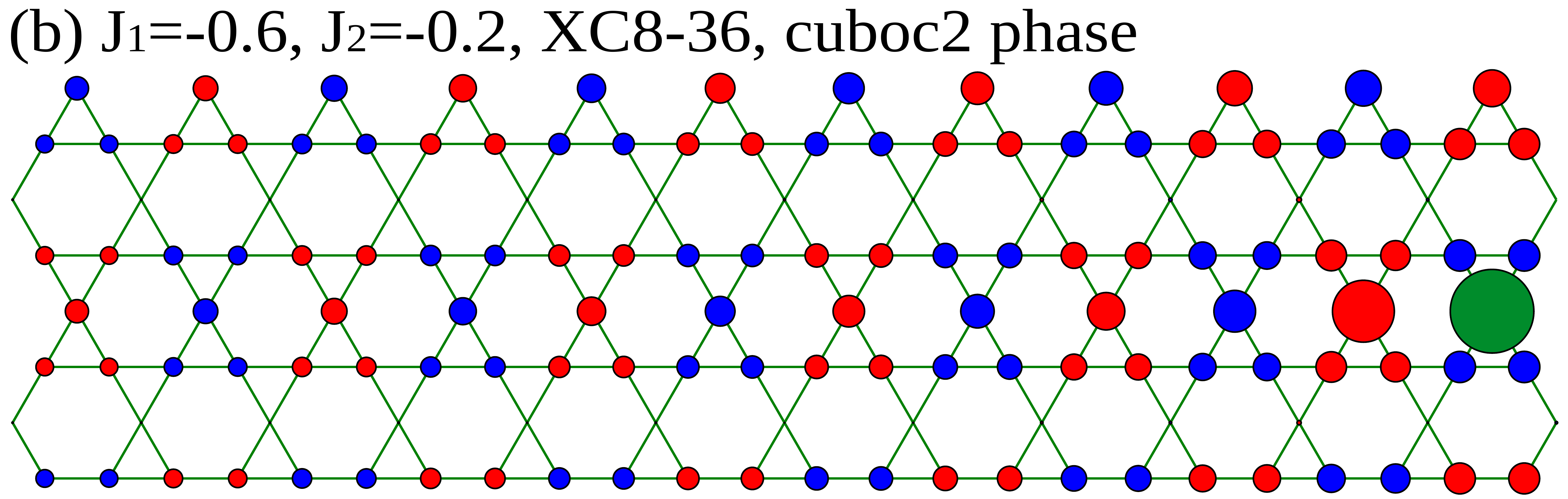}
  \includegraphics[width=1.0\columnwidth]{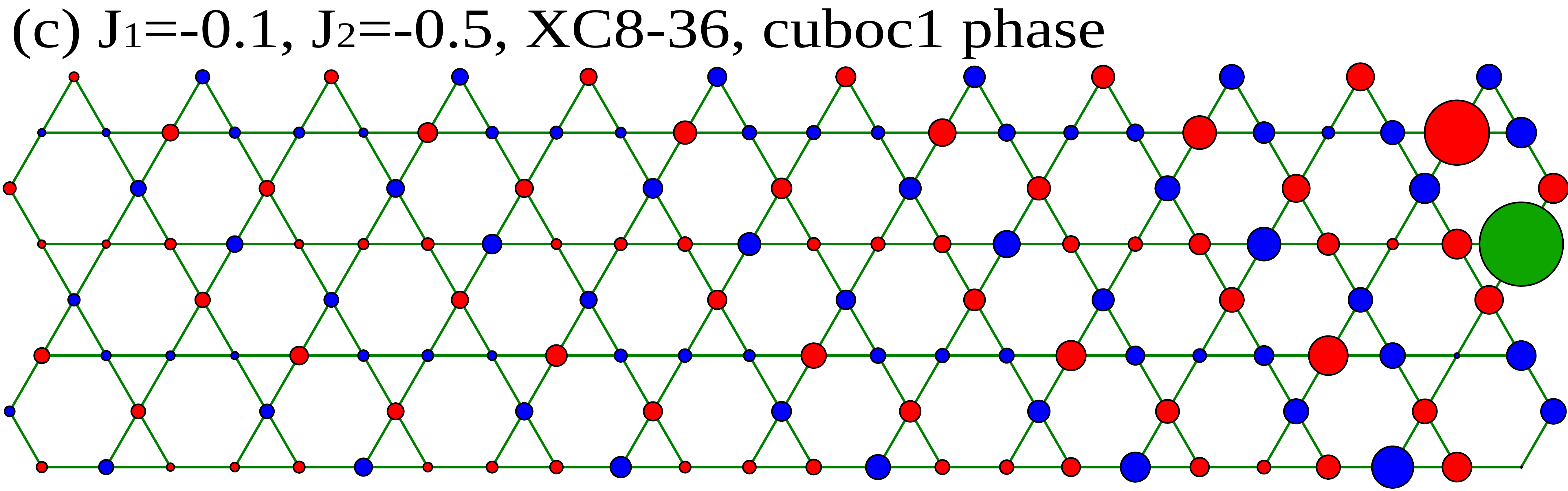}
  \includegraphics[width=1.0\columnwidth]{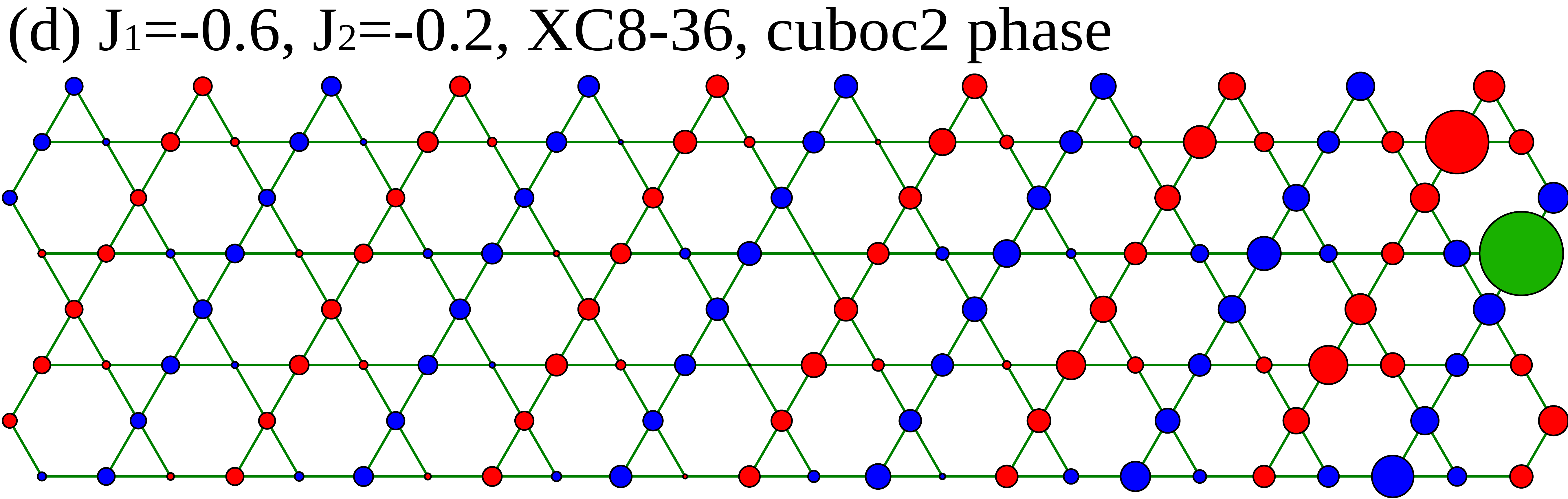}
  \caption{(Color online) Spin correlation functions in real space on
    the XC8-36 cylinders.  Panels  (a) and (b) show the correlations with a
    reference sites on the XC1 chain.  Panels (c) and (d) show the
    correlations with a reference site on the XC3 chain.  The green
    site denotes the reference site in the middle of cylinder. The
    blue and red circles denote the positive and negative
    correlations, respectively. The magnitudes of correlation is
    proportional to the area of the circle.}
  \label{fig:xcspin}
\end{figure}

Fig.~\ref{fig:lattice}(a) shows the geometry of the XC8-8 cylinder,
and the labeling for three types of chains, indicated by dashed lines.
Chains XC2,3 are seen to wind around the cylinder while chains XC1 run
parallel to the cylinder's axis.  In this geometry all chains are
long, i.e. proportional to the cylinder length, which helps to reduce
finite-size effects.  Fig.~\ref{fig:xcspin} shows the real space spin correlations for the XC8-36 cylinder.
When the reference spin, which is shown by a green circle, belongs to chain XC1 (Figs.~\ref{fig:xcspin}(a-b)),
its correlations with the spins from the {\em same} XC1 chain 
are staggered in an antiferromagnetic N\'{e}el pattern. The same is
true for the {\em next-nearest} XC1 chains. 
Note, however, that correlations with the spins from the {\em nearest} XC1 chains are essentially absent.
This is in full agreement with the $[1+(-1)^{\sf{y}+\sf{y'}}]$
structure of the correlations discussed below Eq.~\eqref{eq:os2}.
This in particular is indicative of the cuboc states, in which spins
on successive chains are orthogonal -- see Fig.~\ref{fig:cuboc}.
In addition, correlations between spins on XC1 chain and 
those on the neighboring sites belonging to XC2 chain and XC3 chain
are seen to change sign in going from the $\lambda = 2(J_1 - J_2) > 0$ phase, 
Fig.~\ref{fig:xcspin}(a), to the $\lambda < 0$ one, Fig.~\ref{fig:xcspin}(b).
This too is in agreement with Eq.~\eqref{eq:os2}.
In Figs.~\ref{fig:xcspin}(c-d),
we show the spin correlations with the reference spins on XC3 chain,
which exhibit similar cuboc-like magnetic correlations.

\begin{figure}[h!]
  \includegraphics[width=1.0\columnwidth]{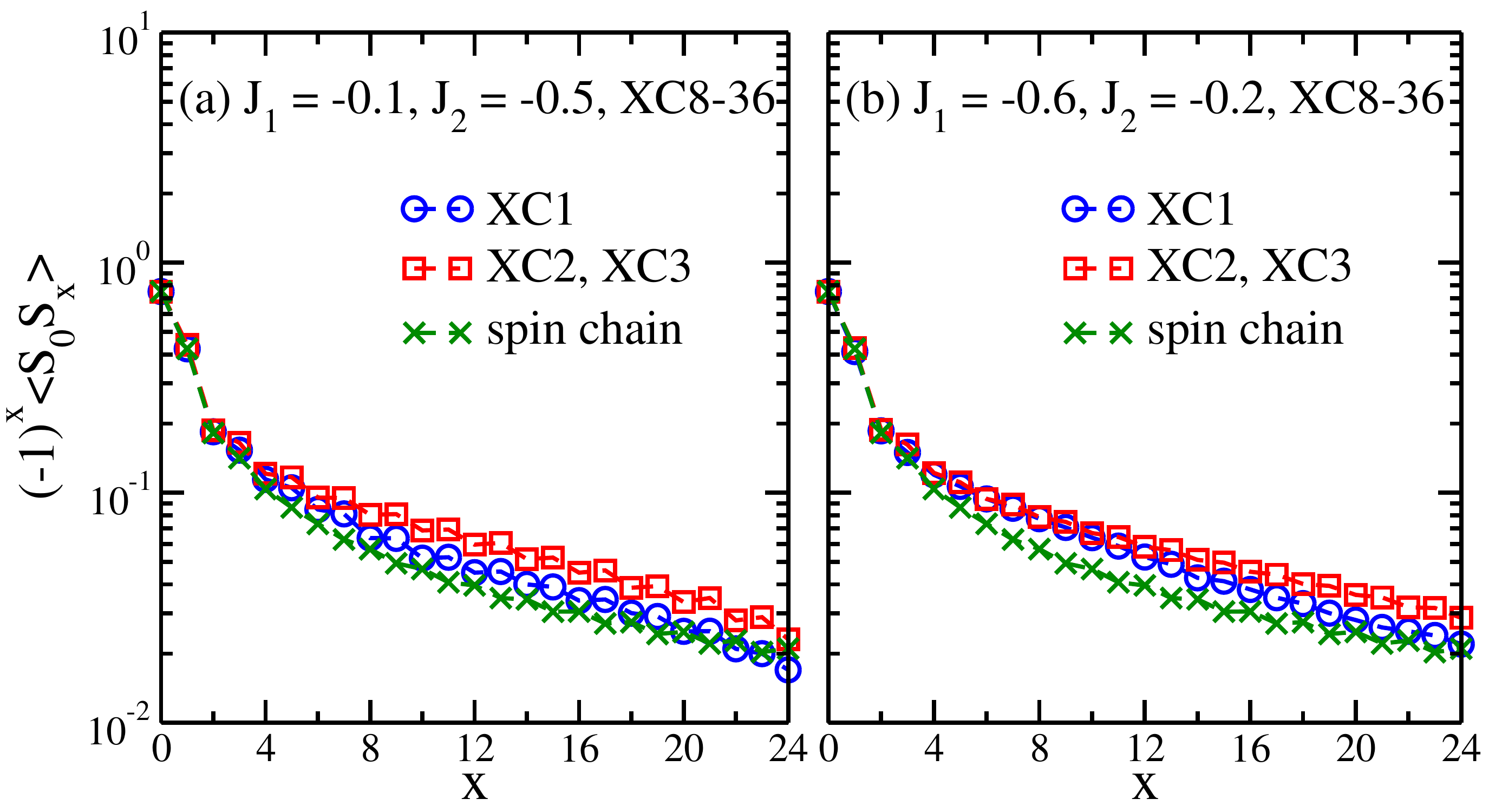}
  \includegraphics[width=1.0\columnwidth]{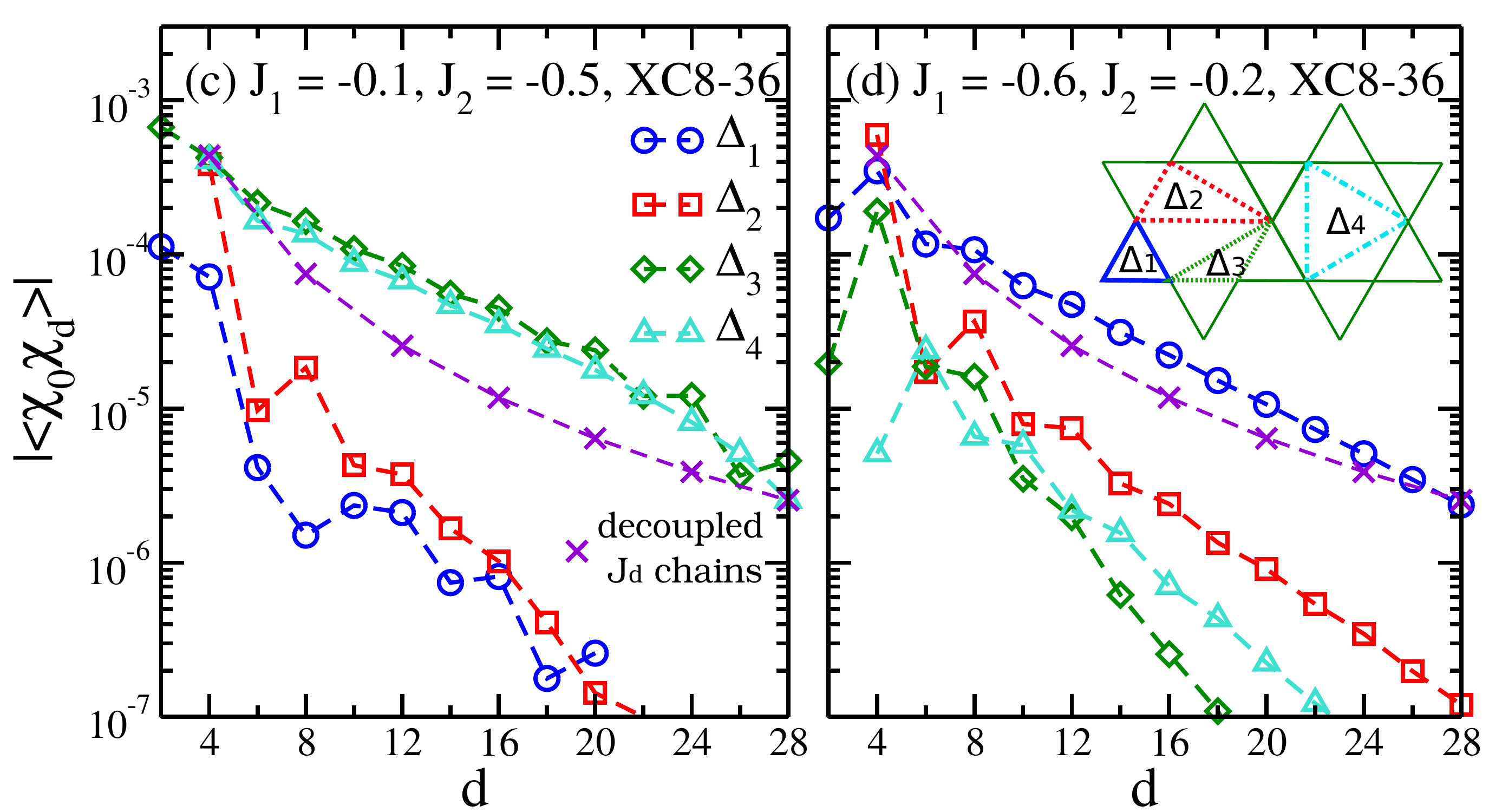}
  \caption{(Color online) Log-linear plot of the correlation functions
    on the XC8-36 cylinder. Panels (a) and (b) give the spin correlations
    along the different $J_d$ chain directions as shown in
    Fig.~\ref{fig:lattice}(a).  Note that the spin correlations are
    enhanced above those of an
    isolated Heisenberg chain, plotted for comparison.  Panels (c) and
    (d) give the chiral
    correlations of different types of triangles, versus the distance
    $d$ along the $x$
    direction.  The  definition of the different triangles is shown in
    the inset of (d).  The triangles which would have
    non-zero chirality in the appropriate classical cuboc state show
    chiral correlations enhanced above those of independent Heisenberg
  chains, while those whose chirality vanishes in the classical state
  have chiral correlations suppressed below the decoupled chain value.}
  \label{fig:xccor}
\end{figure}

To investigate whether the cuboc states have developed long-range
magnetic order, we study the distance dependence of spin and chiral
correlations.  In Figs.~\ref{fig:xccor}(a-b), we plot the spin
correlations along the three $J_d$ chains, as a function of the
distance along the chains, ${\sf x}$, for two systems in the cuboc1
(a) and cuboc2 (b) regimes.  For comparison, the correlations for a
single isolated Heisenberg chain is also shown.  We see that the spin
correlations agree with those of the Heisenberg chain at short
distance ${\sf x} \leq 2$, but are enhanced significantly above them
(note the logarithmic scale) for larger ${\sf x}$.  This is a strong
indication that the system is long-range ordered in the 2d limit.
Despite the enhancement, the spin correlations due continue to decay,
albeit slowly, with distance, rather than saturation.  We attribute
this to the inevitable 2d to 1d crossover which occurs for a
quasi-one-dimensional system.  In fact, for any finite width cylinder,
{\em exponential} decay of the spin correlations is expected at
sufficiently large $L$, due to one-dimensional fluctuations at low
energy -- see Sec.~\ref{sec:quasi-2d-limit}.  The fact that the decay
is relatively weak is a strong indicator that the underlying
two-dimensional state is long-range ordered, rather than a gapless
spin liquid behavior \cite{bieri2014}.

The cuboc states spontaneously break time-reversal symmetry and are
characterized by finite scalar chirality
$\langle \chi_{\bigtriangleup_i} \rangle \neq 0$, where
$\chi_{\bigtriangleup_i} = (\mathbf{S}_{i,1} \times \mathbf{S}_{i,2})
\cdot \mathbf{S}_{i,3}$
and $\mathbf{S}_{i,m}$ ($m=1,2,3$) are the three spins forming
triangle $\bigtriangleup_i (i=1,2,3,4)$ shown in the inset of
Fig.~\ref{fig:xccor}(d).  The distance dependence of the chiral-chiral
correlations
$\langle \chi_{\bigtriangleup_i} \chi_{\bigtriangleup_i} \rangle$ for
each of the four kinds of smallest triangles are plotted in
Figs.~\ref{fig:xccor}(c-d).  

For the cuboc1 state, we expect that the three-spin scalar chiral
order of the triangles $\Delta_{3}, \Delta_{4}$ are non-zero and that
for $\Delta_1,\Delta_2$ vanish, which for the cuboc2 state, we expect
only $\Delta_{1}$ has non-vanishing chirality \cite{messio2011,
  gong2015kagome}. The chiral correlations on the XC8 cylinder are
presented in Figs.~\ref{fig:xccor}(c-d).  Indeed, panel (c) shows that
the chiralities which vanish in the cuboc1 ordered state are extremely
small in the $|J_2|>|J_1|$ region, vanishing rapidly with distance and
taking values close to the precision of the calculation.  The same
holds for the chiralities which are expected to vanish in the cuboc2
state for $|J_1|>|J_2|$, as shown in panel (d).  In either case, the
chiralities which would be expected to be non-zero in the ordered
system still decay exponentially, but are substantially larger.  The
relative magnitudes of the various chiralities are indicative of cuboc
states.  For comparison, we show the chiral correlations of triangle
types 1,3, and 4 for decoupled $J_d$ chains.  One observes that this
lies between than of the strong and weak chiralities in the cuboc
regimes, again indicative of ordered behavior.  However, the apparent
exponential decay of the larger chiralities is {\em not} expected in
the fully quasi-2d limit -- see Sec.~\ref{sec:quasi-2d-limit}. This
indicates that there are still finite size effects due to
insufficiently large $L_y$.  We note that the relatively small
magnitude of the chirality correlations can be understood simply from
the fact that it is a three-spin operator, residing on three different
chains.  Roughly speaking, therefore, the chirality correlations
should have a similar magnitude to the short-distance correlations of
the spins {\em cubed}.  This is generally in accord with the data.

\subsubsection{YC cylinders}
\label{sec:yc-cylinders}

In the YC geometry, two of the chains YC2 and YC3 winding
around the cylinder while the chains of YC1 kind run along the
periodic $y$ direction and are rather short, containing only $4$ or
$6$ sites for YC8 and YC12 cylinders, respectively.  

\begin{figure}[h]
  \includegraphics[width=1.0\columnwidth]{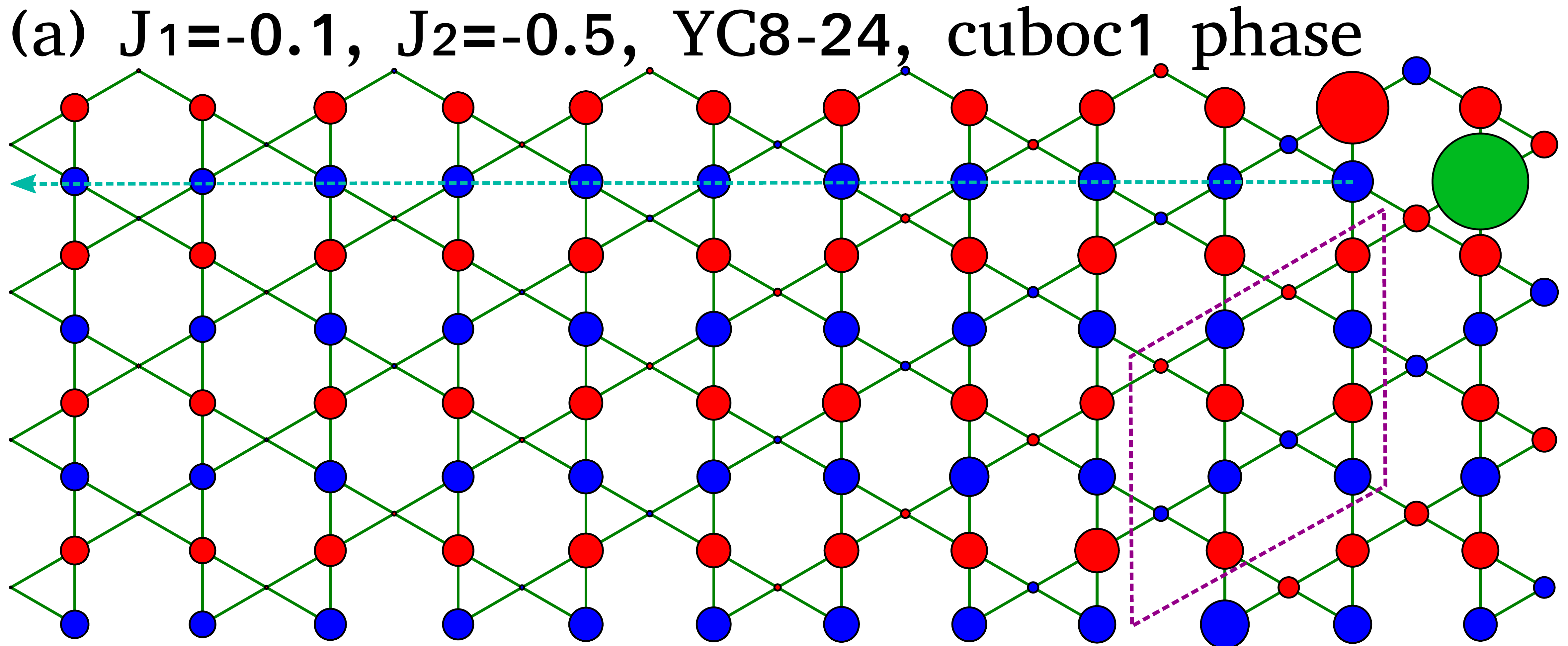}
  \includegraphics[width=1.0\columnwidth]{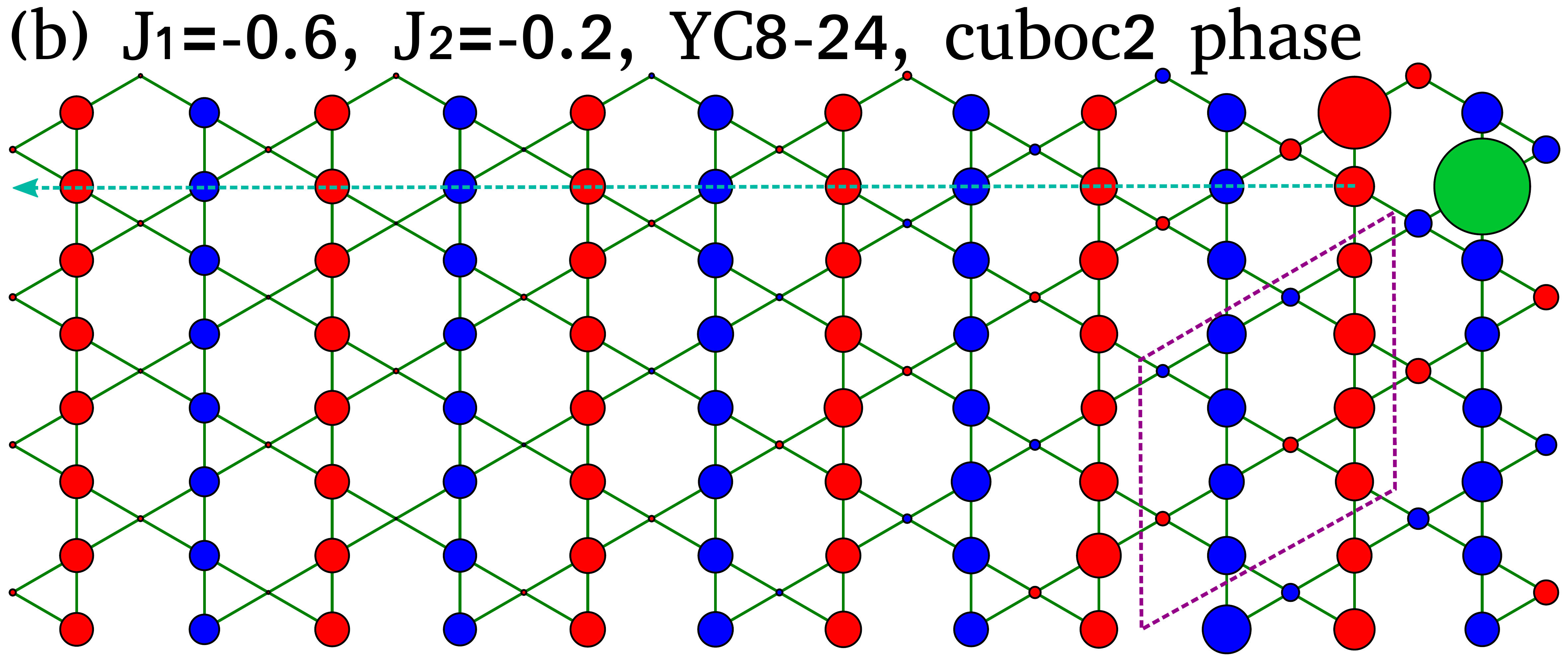}
  \caption{(Color online) Spin correlation functions in real space on
    the YC8-24 cylinders.  The green site is the reference spin in the
    middle of cylinder.  The blue and red circles denote the positive
    and negative correlations, respectively. The magnitudes of
    correlations are proportional to the area of circle.  The dashed
    diamonds denote the $12$-site unit cells.}
  \label{fig:ycspin}
\end{figure}

According to the analytical discussion in
Sec.~\ref{sec:quasi-1d-limit}, the short YC1 chains are strongly
gapped which has the effect of strongly suppressing chiral spin order
and non-coplanar spin correlations associated with it.  In
Fig.~\ref{fig:ycspin} we present the spin correlations with a
reference spin on a YC3 chain. The similarity of the data with the
predicted collinear spin pattern in Fig.~\ref{fig:YCtwosub} is
striking.  The reference spin in Fig.~\ref{fig:ycspin} has strong
correlations with the spins on YC2 and YC3 chains, but very weak
correlations with those on the YC1 chains, consistent with the
proposed large gap formation.  In addition, the observed `striped'
ordering -- ferromagnetic ordering along horizontal/vertical
directions in cuboc1/cuboc2 phases -- is also fully consistent with
simple arguments in Sec.~\ref{sec:quasi-1d-limit}.

\begin{figure}[h!]
  \includegraphics[width=1.0\columnwidth]{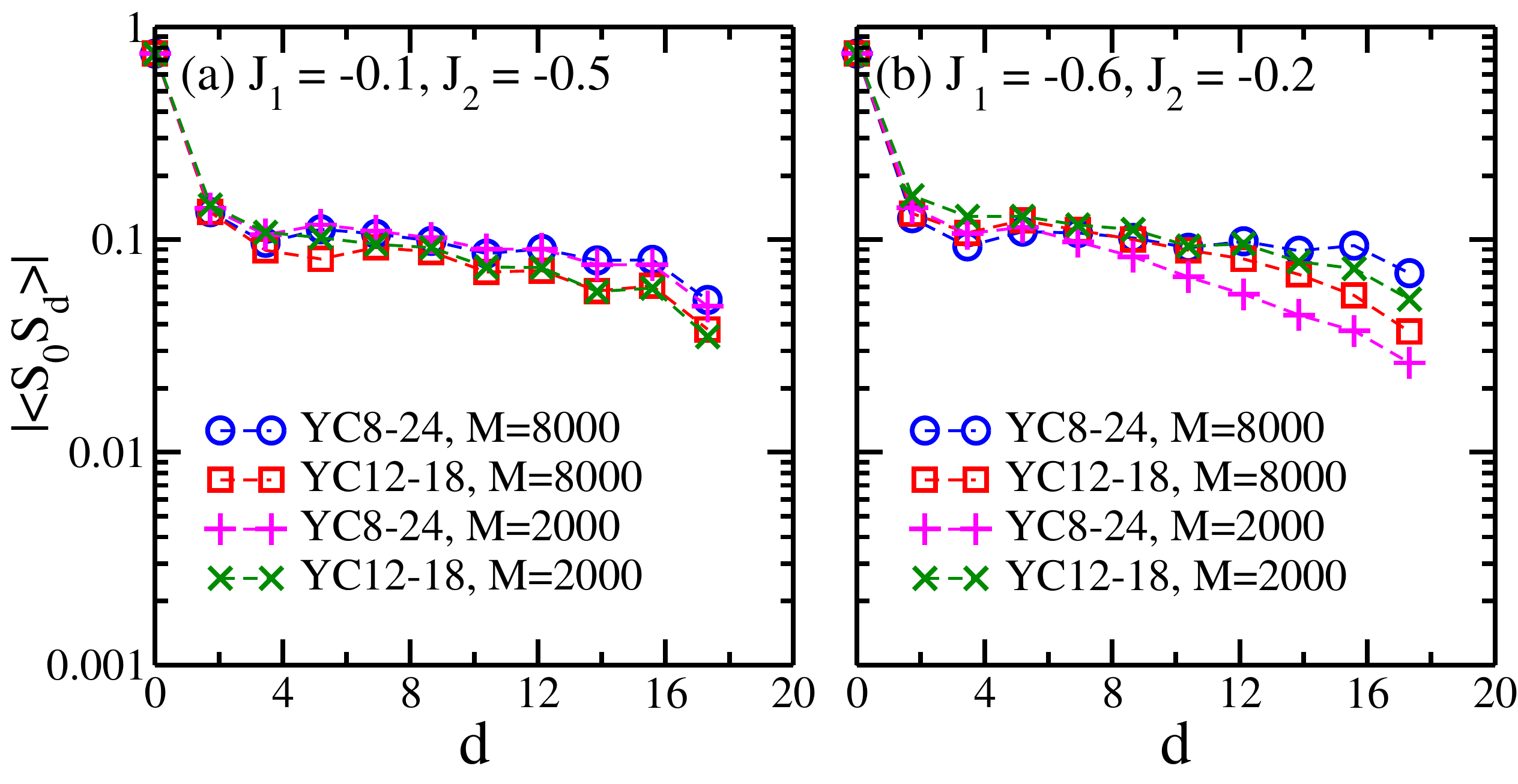}
  \includegraphics[width=1.0\columnwidth]{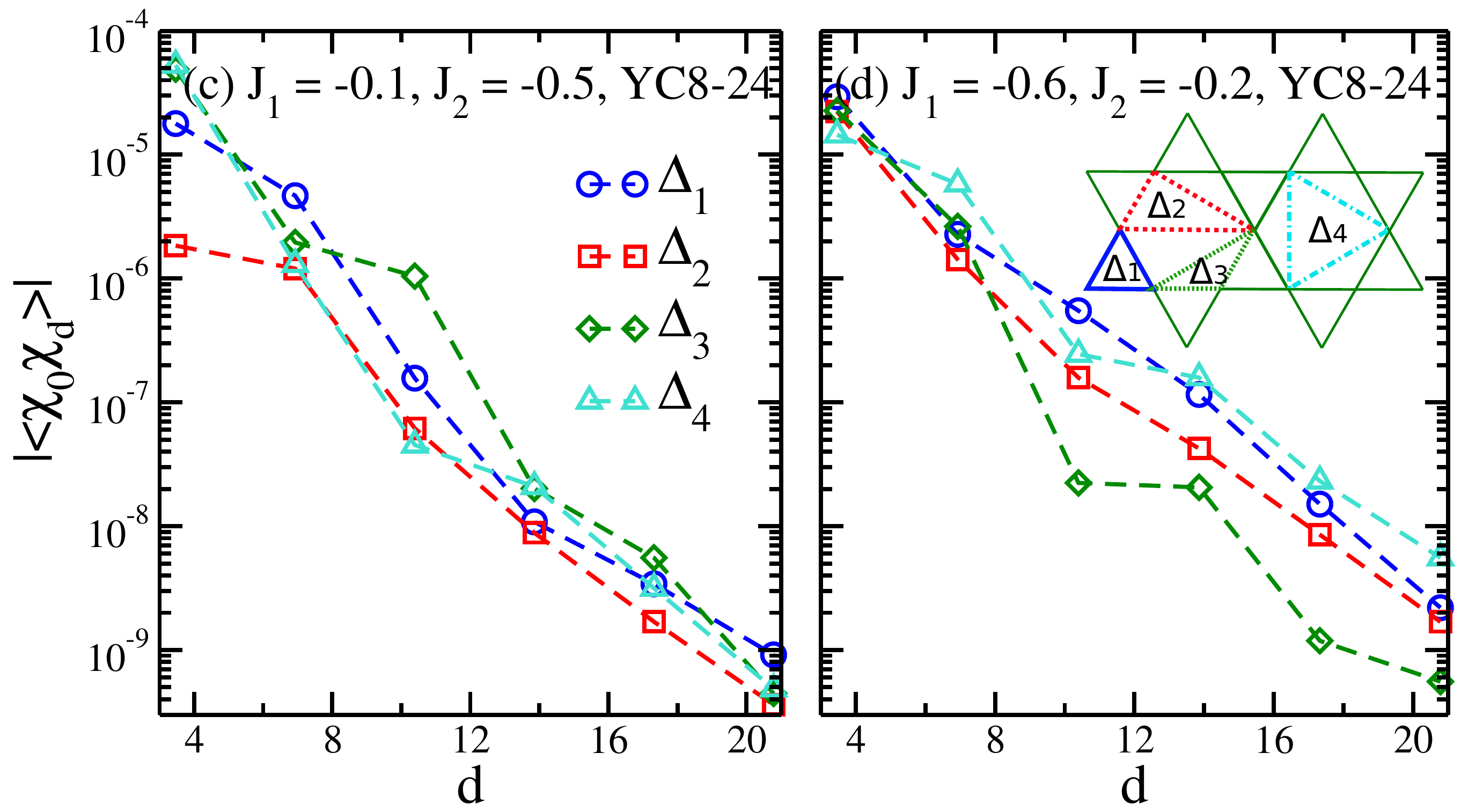}
  \caption{(Color online) Log-linear plots of the correlation
    functions on the YC cylinders.  Panels  (a) and (b) show the spin
    correlations on the YC8-24 and YC12-18 cylinders along 
    the dashed lines in Fig.~\ref{fig:ycspin}.
   Panels (c) and (d) plot the chiral correlations versus the distance
   along the $x$ direction on the YC8-24 cylinder.
  }
  \label{fig:yccor}
\end{figure}

The `gapping out' of the YC1 chains relieves frustration (see
Fig.~\ref{fig:YCtwosub}) and therefore has the effect of enhancing
correlations between spins from the YC2,3 chains. Our data in
Figs.~\ref{fig:yccor}(a-b) reflects this well.  The observed slow decay
of spin correlations with the distance, which is found to be robust
with respect to increase in the number of kept DMRG states, strongly
suggests collinear spin ordering.  Correspondingly, and in agreement
with our numerical findings, we find the chiral correlations to be
strongly suppressed in this geometry, as is shown in
Figs.~\ref{fig:yccor}(c-d).  Together these results vindicate the
conclusion that strong finite size effects on the YC cylinders studied
completely change the ground state from an non-collinear to a
collinear one in the magnetically ordered regions.  We conclude that
the YC cylinders do not give behavior representative of the two
dimensional limit.  However, the consistency with theory shows that we
have an excellent control over finite size effects using the quasi-1d
analytical approach. 

\subsection{VBC phase in the compensated regime}
\label{sec:vbc-phase}

\begin{figure}[h!]
  \includegraphics[width=1.0\columnwidth]{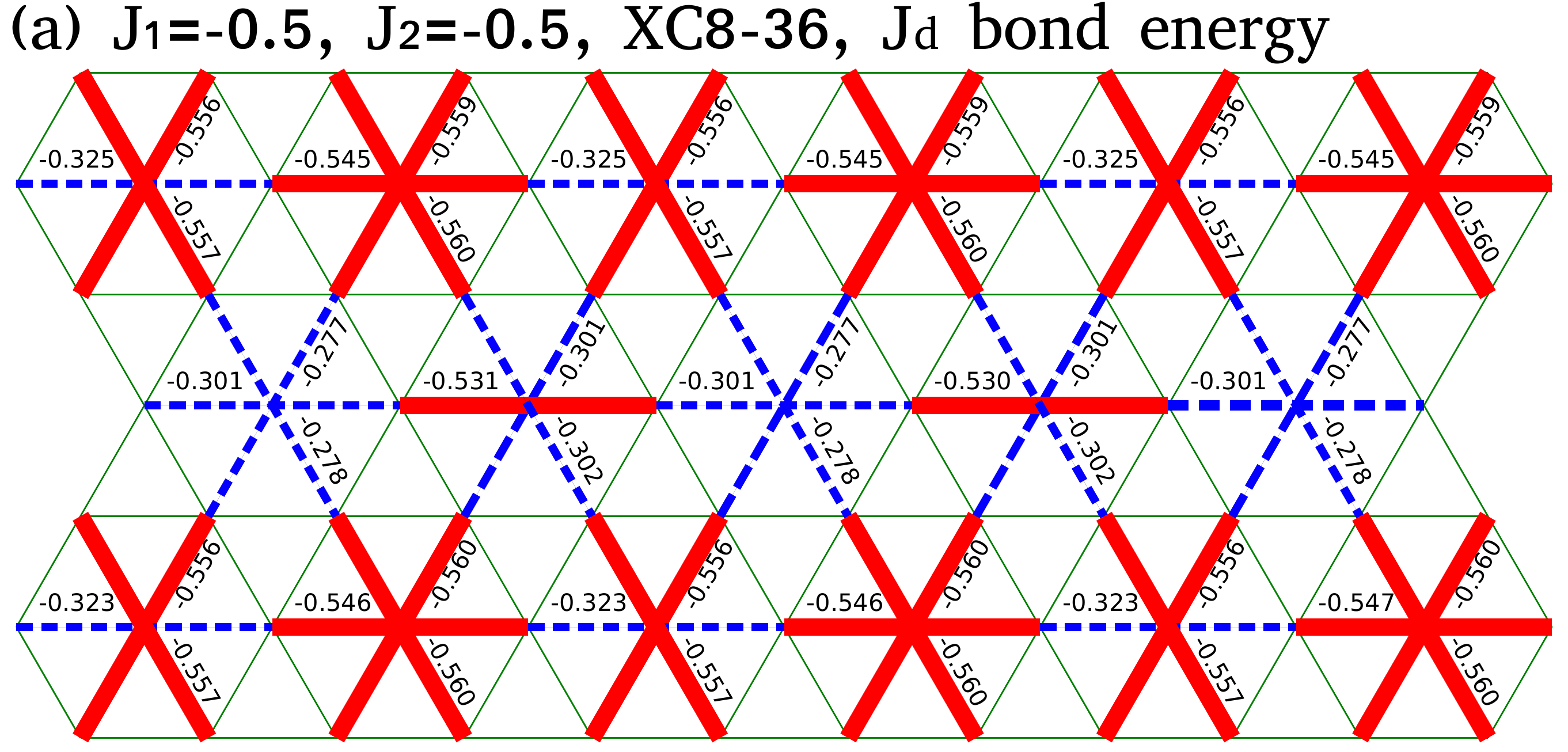}
  \includegraphics[width=1.0\columnwidth]{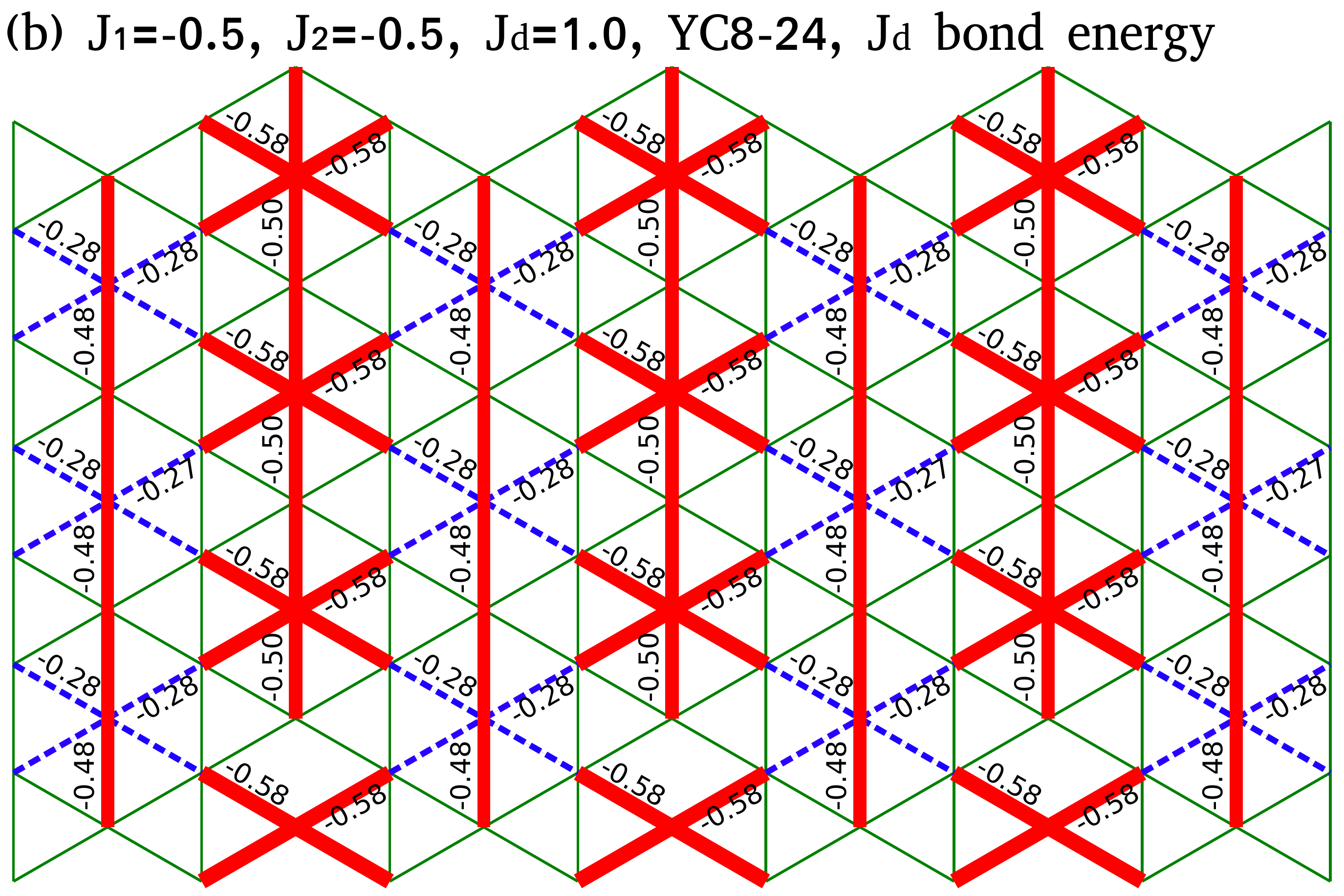}
  \caption{(Color online) VBC pattern for $J_1 = J_2 = -0.5$ on (a)
    XC8-36 and (b) YC8-24 cylinders.  The red solid and blue dashed
    lines denote the strong and weak $J_d$ bond energy
    $\langle {\bm S_i} \cdot {\bm S_j} \rangle$ in the middle of
    cylinder.  Panels (a) and (b) should be compared to the analytical
    predictions in Fig.~\ref{fig:vbc} and Fig.~\ref{fig:ycvbc}.}
  \label{fig:dmrgvbc}
\end{figure}

In the vicinity of the compensated line $J_1 = J_2$, our analytical
studies find a VBC state as shown in Fig.~\ref{fig:vbc} rather than a
decoupled chain state.  To detect the possible lattice translational
symmetry breaking, we have calculated the bond energy pattern on both
XC8 and YC8 cylinders, see Fig.~\ref{fig:dmrgvbc}.  The observed
dimerization pattern indicates an instability of the decoupled chain
state towards a VBC under $J_1, J_2$ perturbations.  On the XC8
cylinder, we find that a dimerization pattern which is very
compatible with the analytical result in two dimensions: compare the
theoretical plot of Fig.~\ref{fig:vbc} with Fig.~\ref{fig:dmrgvbc}(a).  On the YC8 cylinder, the dimerization pattern
also fully agrees with the simpler VBC pattern found in the
short-YC1-chain limit: compare Fig.~\ref{fig:ycvbc} and
Fig.~\ref{fig:dmrgvbc}(b).  The good agreement 
between the DMRG and analytical results indicates that the VBC state
found analytically in Sec.~\ref{sec:compensated-regime} is indeed the
ground state in the compensated regime.

\subsection{Phase boundaries}
\label{sec:dmrg-boundary}

\begin{figure}[h!]
  \includegraphics[width=1.0\columnwidth]{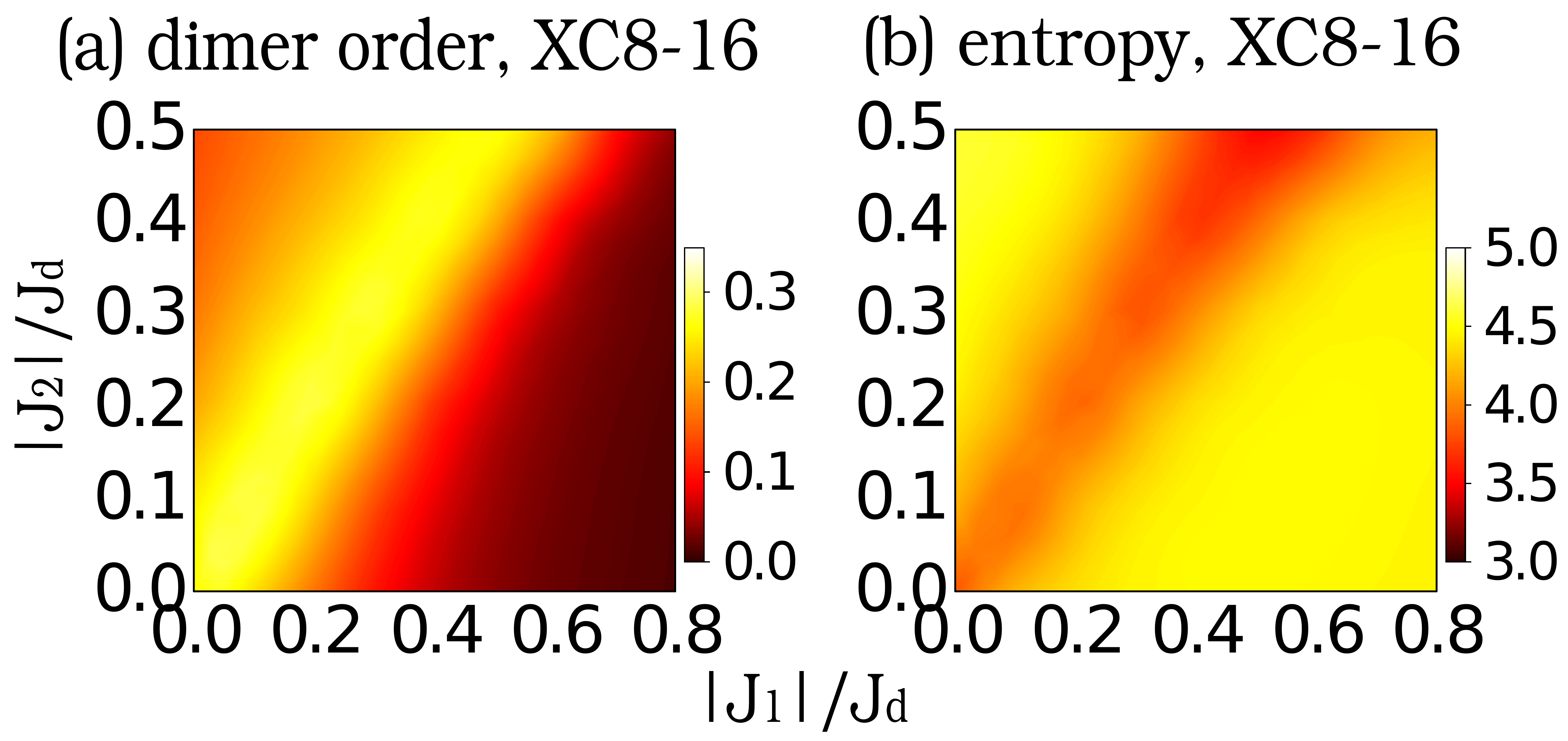}
  \includegraphics[width=1.0\columnwidth]{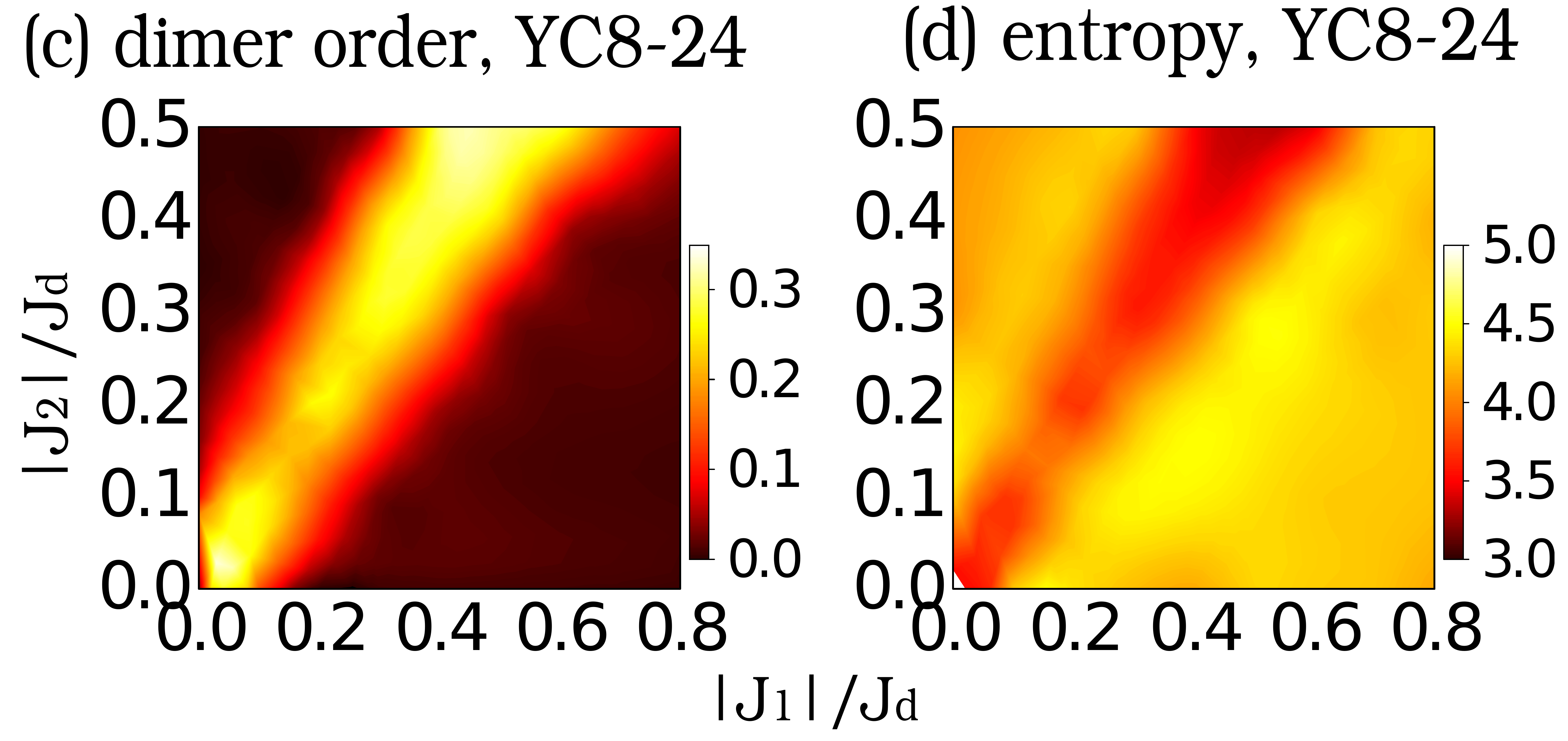}
  \caption{(Color online) Coupling dependence of the bulk dimer order
    parameter and entanglement entropy on the XC8-16 and YC8-24 cylinders. (a) and (c) are
    the $J_d$ bond dimer order parameter, which is defined as the
    difference between the strong and weak bond energy along the $x$
    axis. (b) and (d) are the bipartite entanglement entropy in the bulk of cylinder.}
  \label{fig:boundary}
\end{figure}

In Sec.~\ref{sec:phase-boundary} we argued theoretically that the
phase boundaries between the VBC and the cuboc phases have a
wedge-like shape. To study the phase boundaries numerically, we
calculated the dimer order parameter and the entanglement entropy on
the XC8 and YC8 cylinders as a function of the $J_1, J_2$ couplings.
As shown in Fig.~\ref{fig:boundary}, the $J_d$ bond dimer order
parameter is strongly peaked near the compensated line, indicating the
VBC phase region. At the same time, we find that in the VBC region the
entanglement entropy is strongly suppressed. The approximate phase
boundaries extracted from these two independent quantities are well
correlated with each other.  

The VBC phase region obtained from DMRG
calculations roughly agrees with the wedge-like shape of the
dimerized phase, sketched in Fig.~\ref{fig:phase}, predicted
analytically in the weakly-coupled chain limit. At the same time it is clear that the agreement
is only qualitative as the width of the dimerized region found by DMRG is much wider than
the analytical prediction Eq.~\eqref{eq:32}. We attribute this discrepancy to the well-known fact that
open ends of the spin chain induce finite staggered dimerization which decays slowly, $\propto L^{-1/2}$,
towards the center of the chain of length $L$ \cite{White1993,Tsai2000}. This effect is of course most pronounced
in the weakly-coupled limit $|J_{1,2}| \ll J_d$ (exactly where the discrepancy between numerical and analytical results is largest) where open-ended chain can be best viewed as having 
`pre-formed' dimerization pattern -- the main effect of interchain interactions $J_{1,2}$ is then
to correlate phases of these `pre-formed' patterns between different chains. The very fact that the symmetry of the resulting dimerization 
pattern, Figure~\ref{fig:dmrgvbc}, matches the analytical predictions, Fig.~\ref{fig:vbc} and Fig.~\ref{fig:ycvbc},
implies that the over-estimate of the extend of the VBC region is only a quantitative, and not qualitative, feature of our DMRG study.

\section{Conclusions}

\subsection{Summary}
Motivated by experiments on
the kagom\'{e} antiferromagnet kapellasite \cite{fak2012},
we have studied the spin-$1/2$ $J_1-J_2-J_d$
kagom\'{e} Heisenberg model (with ferromagnetic $J_1,J_2 < 0$, and
antiferromagnetic $J_d > 0$ third-neighbor
coupling across the diagonal of the hexagon) in the $J_d$-dominant
regime by using analytical and Density Matrix Renormalization Group (DMRG) calculations. 
This model has previously been argued to represent a good starting point for this
material \cite{fak2012,janson2008,jeschke2013}.  Both bosonic and
fermionic parton constructions predicted chiral spin liquid phases in
this model, in the regime relevant to kapellasite \cite{bieri2014,fak2012,messio2013time}.  

We approach the problem analytically by considering $J_1$ and $J_2$ as
perturbations to $J_d$, i.e. formally $|J_1,J_2| \ll J_d$, so that the
starting point consists of spin chains formed by strong $J_d$-bonds.
We expect this to work so long as $|J_1|,|J_2|$ remain a fraction of
$J_d$.  Note that this domain of applicability
includes most of the predicted range of the chiral spin liquid phases
found in the most recent parton study \cite{bieri2014}.  We provide a
controlled alternative.  Utilizing a powerful field theory
representation of the one-dimensional chains, we treat the weak
$J_1,J_2$ bonds with the help of the systematic perturbative
renormalization group method and a well-established chain mean field
approximation.

For the parameter regimes away from the compensated line $J_1 = J_2$,
this analysis predicts the non-coplanar cuboc1 and cuboc2 states with
long-ranged magnetic order and finite scalar spin chirality.  In the
compensated regime with $J_1 \simeq J_2$, the leading interchain
interaction vanishes, and quantum fluctuations conspire to generate an
effective four-spin interactions between chains, which is conveniently
expressed in terms of interaction between dimerization densities from
different spin chains. This new interaction promotes an interesting
24-fold degenerate Valence Bond Crystal (VBC) state which breaks
lattice translational and rotational symmetries. We then extend this
analysis to the cylinders of finite circumference and argue that the
XC geometry minimizes finite size effects in comparison to the YC one.

In parallel with this, we carry out unbiased large-scale DMRG
calculations for both the XC and YC cylinder systems. The DMRG results
are summarized in the quantum phase diagram Fig.~\ref{fig:phase},
which agrees well with the analytical predictions. We find cuboc1
($|J_2| > |J_1|$) and cuboc2 ($|J_1| > |J_2|$) magnetic orders with
scalar spin chirality based on the DMRG results on the XC cylinder.
On the YC cylinder, all the chiral correlations decay extremely fast,
in agreement with the analytical prediction that strong finite-size
effects on the YC cylinder of small circumference relieve magnetic
frustration and promote collinear N\'eel-like states over those with
finite scalar chirality.  The two magnetically ordered phases are
separated by a spontaneously dimerized VBC state, whose structure is in excellent
agreement with the analytical predictions.  Like for the magnetically
ordered region, the dimerization pattern is strongly affected by
finite size effects in the YC geometry, but reflects the infinite 2d
limit well in the XC cylinder.

\subsection{Discussion}

Our approach offers unexpectedly deep insight into the kagom\'e kapellasite problem.
It demonstrates -- for the first time, to the best of our knowledge -- emergent 
one-dimensional behavior of a structurally isotropic two-dimensional
problem with hexagonal symmetry. Crucially, this emergent one-dimensionality {\em does not} 
imply spontaneous separation of the model into a collection of decoupled spin chains.
Rather, it offers a valuable insight into separation of relevant energy scales in the problem.
At the highest energy $\propto J_d$, the spin chains are real -- spin fluctuations,
as probed for example by the dynamic spin structure factor in inelastic neutron scattering experiments, have
a strong one-dimensional character and mostly propagate along one of the three available chain directions.
This interesting feature is, however, strongly masked by the hexagonal symmetry of the lattice:  
to the untrained eye 
excitations running at $\pm 120^\circ$ to each other will probably appear as almost isotropic two-dimensional modes. 
Even more importantly, at this high energy spin excitations are
{\em fractionalized} -- they are the spin-1/2 spinons of the emergent spin chains. This immediately implies
that in inelastic neutron scattering experiments they show up as a broad multi-particle continuum 
extending up to energy $\propto \pi J_d$. Kinematic effects, of the kind described previously for
structurally-anisotropic antiferromagnet Cs$_2$CuCl$_4$ \cite{kohno2007}, 
may produce coherent spin-1 triplon excitations in some parts of the Brillouin zone.

At much lower energy $\propto J_{1,2}$, the two-dimensional cuboc order sets in. One-dimensional spinons
bind into spin-1 spin waves which propagate truly isotropically in the kagom\'e lattice.  
The energy-dependent evolution of spinons into spin waves is quite complex and represents
a challenging open theoretical problem, beyond the scope of this work,
and relevant to many systems. 
Interestingly, in the vicinity of the compensated line $J_1 \approx J_2$ the characteristic energy scale
for the two-dimensional cross-over is yet smaller, $\propto J_{1,2}^2/J_d \ll J_{1,2}$. In this VBC regime
all spin excitations are gapped.  

This brief description makes it clear that dynamic response of the model Hamiltonian \eqref{hamiltonian}
is very complex. It is expected to show a number of one-dimensional features,
such as a broad incoherent continuum and strong dispersion along the three crystallographic chain directions.
This, we insist, does not imply a spin-liquid ground state. 
Our extensive analytical and DMRG calculations find no evidence in support of
the previously suggested \cite{bieri2014} gapless chiral spin liquid
states for $J_1 \neq J_2$ as well as of the decoupled chain state in the neighborhood of $J_1 = J_2$ line.  

Turning now to the real material kapellasite, we observe that in the regime
of parameters relevant to it \cite{janson2008, jeschke2013,bernu2013}, 
the ground state, according to our phase diagram Fig.~\ref{fig:phase},
is the magnetically ordered cuboc2 phase rather than a spin liquid state.
The experimentally observed spin liquid behavior \cite{ker2014,bernu2013,fak2012} can
be interpreted in two different ways. The first consists in the assumption that
observed `spin-liquid' features are remnants of the emergent high-energy spinons 
described above. The alternative explanation points out the importance of disorder
which, according to recent NMR experiments, reaches a very large level -- 
up to $\sim$ 25\% of spins are missing from the kagom\'e planes \cite{ker2014}.

Interestingly, the `one-dimensional' framework proposed here can be
straightforwardly applied to the analysis of disorder effects as
well. The effect of non-magnetic disorder on a spin-1/2 chain is well
understood~\cite{eggert1992} and its
experimental manifestations in neutron scattering experiments have
recently been identified in Ref.~\onlinecite{zheludev2013}.  It therefore seems that
extending our `one-dimensional' perspective to the kapellasite model
with disorder is within reach. We leave such studies to the future.

Finally, the fact that established phase diagram does not include the
previously suggested spin-liquid phase demonstrates the limitations
of the frequently used parton mean-field approach, even when improved
by Gutzwiller projection. It also shows that the accepted minimal kapellasite 
model \eqref{hamiltonian} is not general enough to provide a new path to 
spin-liquid phases of magnetic matter.

\acknowledgments
We acknowledge discussions with S.~Bieri and C.~Lhuillier.
This research is supported by the state of Florida (S.S.G.), 
National Science Foundation Grants DMR-1157490 (S.S.G. and K.Y.), 
PREM DMR-1205734 (W.Z.), DMR-1442366 (K.Y.), DMR-1507054 (O.A.S.), 
DMR-1408560 (D.N.S), and DMR-1506119 (L.B.). We also acknowledge
partial support from NSF Grant DMR-1532249 for computational resource.

\bibliographystyle{apsrev}
\bibliography{kap-new}{}

\clearpage

\appendix

\section{Symmetries and transformations}
\label{sec:symm-transf}

Here we discuss the symmetries of the kagom\'e lattice.  A sufficient
set of generators for the full space group consists of two elementary
translations, a ${\mathcal C}_6$ rotation about the center of a hexagon, and a
reflection ${\mathcal P}$ through a line passing through a site and a hexagon
center.   We begin with the description of these operations in terms
of the primitive vectors ${\bm a}_i$ given in the main text.  Under
the translations, we have
\begin{eqnarray}
  \label{eq:24}
  {\mathcal T}_1: && {\bm x} \rightarrow {\bm x} + {\bm a}_1, \nonumber \\
  {\mathcal T}_2:&& {\bm x} \rightarrow {\bm x} + {\bm a}_2.
\end{eqnarray}
The rotation and reflection act according to
\begin{eqnarray}
  \label{eq:25}
  {\mathcal C}_6: && {\bm a}_1 \rightarrow - {\bm a}_{q-1}, \nonumber \\
  {\mathcal P}: && {\bm a}_1 \leftrightarrow {\bm a}_2, \qquad {\bm a}_3
              \rightarrow {\bm a}_3.
\end{eqnarray}
From these definitions, we can work out the action of these operations
in the chain basis, in which a site is represented in the form
$(q,{\sf x}, {\sf y})$.  Using the definitions of the sites, ${\bm x}=
({\sf x}+\frac{1}{2}){\bm a}_q + {\sf y} {\bm a}_{q+1}$, and keeping
in mind the relation $\sum_q {\bm a}_q = 0$, we obtain for the
translations
\begin{eqnarray}
  \label{eq:26}
  {\mathcal T}_1: & \left\{
                     \begin{array}{lcl}
                       (1,{\sf x}, {\sf y}) & \rightarrow & (1,{\sf
                                                            x}+1,{\sf
                                                            y}) \\
                       (2,{\sf x}, {\sf y}) & \rightarrow & (2,{\sf
                                                            x}-1,{\sf
                                                            y}-1) \\
                       (3,{\sf x}, {\sf y}) & \rightarrow & (3,{\sf
                                                            x},{\sf
                                                            y}+1)
                     \end{array} \right., \nonumber \\
  {\mathcal T}_2: & \left\{
                     \begin{array}{lcl}
                       (1,{\sf x}, {\sf y}) & \rightarrow & (1,{\sf
                                                            x},{\sf
                                                            y}+1) \\
                       (2,{\sf x}, {\sf y}) & \rightarrow & (2,{\sf
                                                            x}+1,{\sf
                                                            y}) \\
                       (3,{\sf x}, {\sf y}) & \rightarrow & (3,{\sf
                                                            x}-1,{\sf
                                                            y}-1)
                     \end{array} \right. .
\end{eqnarray}
Under the point group operations we obtain
\begin{eqnarray}
  \label{eq:27}
    \mathcal{C}_6: & (q,{\sf x}, {\sf y}) \rightarrow (q-1,-{\sf
                      x}-1,-{\sf y}), \nonumber \\
  \mathcal{P}: & \left\{
                     \begin{array}{lcl}
                       (1,{\sf x}, {\sf y}) & \rightarrow & (2,{\sf
                                                            x}-{\sf y},-{\sf y}) \\
                       (2,{\sf x}, {\sf y}) & \rightarrow &(1,{\sf
                                                            x}-{\sf y},-{\sf y}) \\
                       (3,{\sf x}, {\sf y}) & \rightarrow &(3,{\sf
                                                            x}-{\sf y},-{\sf y})
                     \end{array} \right. .
\end{eqnarray}
Now with this in hand, we can evaluate the transformations of the
dimerization operators, $\varepsilon_{q,{\sf y}}  = \sum_{\sf x}
(-1)^{\sf x} {\bm S}_{q,{\sf y}}({\sf x}) \cdot  {\bm S}_{q,{\sf
    y}}({\sf x}+1)$.  We find
\begin{eqnarray}
  \label{eq:28}
   {\mathcal T}_1: & \left\{
                     \begin{array}{lcl}
                       \varepsilon_{1,{\sf y}} & \rightarrow & -
                                                               \varepsilon_{1,{\sf
                                                               y}} \\
                       \varepsilon_{2,{\sf y}} & \rightarrow & -
                                                               \varepsilon_{2,{\sf
                                                               y}-1}
                       \\
                       \varepsilon_{3,{\sf y}} & \rightarrow &
                                                               \varepsilon_{3,{\sf
                                                               y}+1}
                     \end{array} \right., \nonumber \\
  {\mathcal T}_2: & \left\{
                     \begin{array}{lcl}
                     \varepsilon_{1,{\sf y}} & \rightarrow &
                                                               \varepsilon_{1,{\sf
                                                               y}+1} \\
                       \varepsilon_{2,{\sf y}} & \rightarrow & -
                                                               \varepsilon_{2,{\sf
                                                               y}}
                       \\
                       \varepsilon_{3,{\sf y}} & \rightarrow &
                                                              - \varepsilon_{3,{\sf
                                                               y}-1}
                     \end{array} \right. .
\end{eqnarray}
and
\begin{eqnarray}
  \label{eq:29}
      \mathcal{C}_6: & \varepsilon_{q,{\sf y}} \rightarrow
                       \varepsilon_{q-1,-{\sf y}} \nonumber \\
  \mathcal{P}: & \left\{
                     \begin{array}{lcl}
                       \varepsilon_{1,{\sf y}} & \rightarrow &
                                                              (-1)^{\sf
                                                               y}\varepsilon_{2,-{\sf y}} \\
                       \varepsilon_{2,{\sf y}} & \rightarrow &
                                                              (-1)^{\sf
                                                               y}\varepsilon_{1,-{\sf
                                                               y}} \\
                       \varepsilon_{3,{\sf y}} & \rightarrow &
                                                              (-1)^{\sf
                                                               y}\varepsilon_{3,-{\sf y}}
                     \end{array} \right. .
\end{eqnarray}

Finally, we can use this to give the transformation properties for
$W_q$ and $V_q$.  Under translations,
\begin{eqnarray}
  \label{eq:30}
  \mathcal{T}_1: & \begin{pmatrix} W_1 \\ W_2 \\
      W_3 \end{pmatrix} \rightarrow \begin{pmatrix} - W_1 \\ - W_2 \\
      W_3 \end{pmatrix}, \qquad  \begin{pmatrix} V_1 \\ V_2 \\
      V_3 \end{pmatrix} \rightarrow \begin{pmatrix} - V_1 \\ V_2 \\
      -V_3 \end{pmatrix}\nonumber \\
 \mathcal{T}_2: &  \begin{pmatrix} W_1 \\ W_2 \\
      W_3 \end{pmatrix} \rightarrow \begin{pmatrix}  W_1 \\ - W_2 \\
      -W_3 \end{pmatrix}, \qquad  \begin{pmatrix} V_1 \\ V_2 \\
      V_3 \end{pmatrix} \rightarrow \begin{pmatrix} - V_1 \\ -V_2 \\
      V_3 \end{pmatrix},
\end{eqnarray}
and under the point operations,
\begin{eqnarray}
  \label{eq:31}
  \mathcal{C}_6: &  W_q \rightarrow W_{q-1}, \qquad V_q \rightarrow
                   V_{q-1}, \nonumber \\
  \mathcal{P}: & \left\{ \begin{array}{ll}
                           W_1 \rightarrow V_2, \qquad & V_1
                                                         \rightarrow
                                                         W_2 \\
                           W_2 \rightarrow V_1, \qquad & V_2
                                                         \rightarrow
                                                         W_1 \\
                           W_3 \rightarrow V_3, \qquad & V_3
                                                         \rightarrow
                                                         W_3
                         \end{array} \right. .
\end{eqnarray}

\end{document}